\documentclass[12pt]{article}

\usepackage[english]{babel}
\usepackage[T1]{fontenc}
\usepackage[utf8]{inputenc}
\usepackage{amssymb,amsmath,array,hhline}
\usepackage{latexsym}
\usepackage{graphicx}
\usepackage[final]{pdfpages}
\usepackage{verbatim}

\title{Topological invariants of edge states \\ for periodic two-dimensional models}

\author{Julio Cesar Avila$^1$, Hermann Schulz-Baldes$^{1,2}$, Carlos Villegas-Blas$^1$
\\
\\
{\small $^1$ Instituto de Matematicas, Cuernavaca, UNAM, Mexico}
\\
{\small $^2$ Department Mathematik, Universit\"at Erlangen-N\"urnberg, Germany}
}

\date{ }

\newtheorem{theo}{Theorem}
\newtheorem{defini}{Definition}
\newtheorem{proposi}{Proposition}

\newcommand{\CM}{{\mathbb C}}
\newcommand{\NM}{{\mathbb N}}
\newcommand{\RM}{{\mathbb R}}
\newcommand{\SM}{{\mathbb S}}
\newcommand{\TM}{{\mathbb T}}
\newcommand{\ZM}{{\mathbb Z}}

\newcommand{\Ee}{{\cal E}}

\newcommand{\Ss}{{\cal S}}
\newcommand{\Oo}{{\cal O}}
\newcommand{\Tr}{\mbox{\rm Tr}}
\newcommand{\Tt}{{\cal T}}

\newcommand{\Nn}{{\cal N}}

\newcommand{\Jj}{{\cal J}}

\newcommand{\Hh}{{\cal H}}

\newcommand{\one}{{\bf 1}}

\newcommand{\Uu}{\mathcal{U}}
\newcommand{\Bb}{\mathcal{B}}
\newcommand{\Ch}{\mbox{\rm Ch}}
\newcommand{\SCH}{\mbox{\rm SCh}}
\newcommand{\EI}{\mbox{\rm Ei}}
\newcommand{\SEI}{\mbox{\rm SEi}}
\DeclareMathOperator{\e}{e}
\DeclareMathOperator{\vol}{Vol}

\begin{document}

\maketitle

\begin{abstract}
Transfer matrix methods and intersection theory are used to calculate the bands of edge states for a wide class of periodic two-dimensional tight-binding models including a sublattice and spin degree of  freedom. This allows to define topological invariants by considering the associated Bott-Maslov indices which can be easily calculated numerically. For time-reversal symmetric systems in the symplectic universality class this leads to a $\ZM_2$-invariant for the edge states. It is shown that the edge state invariants are related to Chern numbers of the bulk systems and also to (spin) edge currents, in the spirit of the theory of topological insulators.

\vspace{.1cm}

MSC: 81V70, 19L10, 82B20
\end{abstract}

\vspace{.5cm}

\section{Introduction}
\label{sec-intro}

Edge or surface states for models of solid state physics have been studied since the early contributions of Tamm \cite{Tam} and Shockley \cite{Sho} in the 1930's. More recently, bands of edge states played a prominent role in the theory of the quantum Hall effect \cite{Hal,FST}. A calculation of edge states of tight-binding models using transfer matrix methods seems to have been done for the first time by Hatsugai \cite{Hat}. He considered a square lattice with rational magnetic flux (Harper model) and determined the edge states by studying the contracting direction of the $2\times 2$ transfer matrices after partial Fourier transform in the direction along the boundary. This provides bands of such states indexed by the corresponding quasi-momentum. These bands typically carry edge currents. Later on, this approach was also applied to study edge states of graphene in a magnetic field \cite{HFA}. Moreover, this allows to define a topological invariant of the edge state bands which is then shown to be connected to Chern numbers of the system without edge, a fact that is crucial for the quantum Hall effect. Both the edge invariant and its connection to bulk invariants have been generalized to disordered systems in \cite{SKR,KRS}.  Compactly supported edge states for zig-zag boundaries of graphene \cite{FWNK} are of different type. They lead to flat bands of edge states which cannot carry edge currents. This type of compactly supported edge states has been thoroughly analyzed in \cite{NGK}. 

\vspace{.2cm}

The first goal of this paper is to present a conceptual approach for the calculation of edge states when the transfer matrices are of larger size. This naturally appears when considering systems with a spin degree of freedom (not necessarily conserved) and when the lattice has a sublattice structure such as for the honeycomb lattice. It appears that exact diagonalization of finite volume operators has been the only recourse in these situations to calculate the edge states in the physics literature. In the present paper the edge states are determined by studying the intersection of the Lagrangian plane of boundary conditions with the Lagrangian plane of decaying directions of the hyperbolic transfer matrix for energies outside of the band of the periodic operator. This intersection theory (see Theorem~\ref{theo-edgecalc}) provides a tool to easily calculate the edge states analytically and numerically. The main ingredient of proof is the analysis of bound states for Jacobi matrices with matrix entries, as developed in the appendix. 

\vspace{.2cm}

The intersection number approach furthermore leads naturally to the second goal of the paper, namely to define and analyze topological invariants of the edge state bands. In fact, at fixed energy this edge index $\EI(E)$ at energy $E$ is simply the Bott-Maslov index associated to the periodic path (parametrized by quasi-momentum) of Lagrangian planes given by the contracting directions of the transfer matrix. It can also be shown that the edge index is equal to the edge Hall conductance as defined in \cite{Hat,SKR}. This also allows to calculate the edge index as a winding number of the unitaries associated to these Lagrangian planes. In the spirit of \cite{Hat,KRS,EG} one can, moreover, prove the simple relation $\Ch(P)=\EI(E_+)-\EI(E_-)$ between the Chern number of the spectral projection $P$ of the planar Hamiltonian on a band lying in the interval $(E_-,E_+)$ and the edge indices in the neighboring gaps (see Theorem~\ref{theo-edgechannelcount}). Because it is numerically so simple to calculate the edge indices, this also provides an efficient means to calculate Chern numbers. It is also shown how edge indices and Chern numbers can be refined to spin edge indices and spin Chern numbers for systems with one conserved spin component (at least up to a small perturbation \cite{Pro}) and that still the same correspondence holds, as well as a connection to spin edge currents (Theorem~\ref{theo-spincurrentquant}).

\vspace{.2cm}

The third main goal concerns systems with time-reversal symmetry with particular focus on the symplectic universality class. For such systems the edge indices and Chern numbers all vanish. As first exhibited in the concrete situation of a honeycomb lattice with spin-orbit interaction by  Kane-Mele \cite{KM}, one can nevertheless associate a $\ZM_2$-index to these systems which allows to distinguish trivial from non-trivial topology of the Bloch bands. The Kane-Mele $\ZM_2$-invariant is defined using the vorticity of a certain Pfaffian, but another way to define a $\ZM_2$-invariant is to consider the spin Chern numbers modulo $2$ \cite{SWSH,Pro}, see details in Section~\ref{sec-SpinChern}. Unfortunately there does not seem to be a proof available that these two $\ZM_2$-indices are equal. Both are associated to the planar models and do not infer edge states. Here we define a new $\ZM_2$-index associated to the bands of edge states (see Section~\ref{sec-edgeZ2inv}). Roughly, it is the Bott-Maslov index modulo 2, but we also provide a winding number calculation for this invariant. Again it is simple to calculate it numerically. Furthermore, there is a relation of the edge $\ZM_2$-index to the Chern $\ZM_2$-index (see Theorem~\ref{theo-edgechannelcount2}) so that the latter can again be easily deduced from the edge indices. Finally, it is proved in Theorem~\ref{theo-spinedgecur} that a non-trivial $\ZM_2$-invariant implies that the spin edge currents do not vanish (albeit they are not quantized). This is in agreement with the stability of the absolutely continuous spectrum proved for a toy model for edge states \cite{SS}, and also in line with the general theory of topological insulators.

\vspace{.2cm}

Apart from these structural general results, the paper also contains a Section~\ref{sec-examples} which describes concrete models entering into the framework presented in Section~\ref{sec-models}. A transfer matrix formalism is, in particular, built up for the Ando model (spin-orbit coupling), operators on the triangular and hexagonal lattice, as well as the Kane-Mele model including spin-orbit coupling and a Rashba term. One reason to do this in such great detail is that disorder is easily added to these periodic models and that there are numerous interesting questions related to them, so that we hope our presentation to be helpful for further studies. In order to exemplify the general theory of edge states and invariants, the edge invariants are calculated numerically for the Harper model and the Kane-Mele model by using a short mathematica code.

\vspace{.2cm}

\noindent {\bf Acknowledgements:} We thank CONACYT and PAPIIT-UNAM IN109610-2 as well as the DFG for financial support. A few months after this work was completed, Graf and Porta posted a paper \cite{GP} which, beneath other things, also extends some of the results of the present work. 
 
\section{Two-dimensional translation invariant Hamiltonians}
\label{sec-models}

This paper is about a quantum particle with spin $s\in\NM/2$ on the square lattice $\ZM^2$ where over each point of $\ZM^2$ there are $R\in\NM$ internal degrees of freedom. Hence the Hilbert space is $\Hh=\ell^2(\ZM^2)\otimes\CM^R\otimes \CM^r$ where $r=2s+1$. In the examples given in Section~\ref{sec-examples}, it is shown that besides operators on the square lattice, also operators on the triangular lattice and the honeycomb lattice can be described by using $R\geq 2$ and next to next nearest neighbor hopping terms. For convenience, let us also set $L=Rr$ and write $\CM^L=\CM^R\otimes \CM^r$. 

\vspace{.2cm}

On the Hilbert space $\Hh$ act the standard shift operators $S_1$ and $S_2$ in the two directions defined by $(S_1\phi)_{n_1,n_2}=\phi_{n_1-1,n_2}$ and $(S_2\phi)_{n_1,n_2}=\phi_{n_1,n_2-1}$. They commute and a magnetic field in a particular gauge can be introduced via the coefficient matrices in \eqref{eq-genH}. Let us also introduce $S_3=S_1^*S_2$ describing the next to next nearest neighbor hopping terms. We consider translation invariant Hamiltonians of the form
\begin{equation}
\label{eq-genH}
H\;=\;\sum_{i=1,2,3}(T_i^*S_i+T_iS_i^*)\;+\;V
\;,
\end{equation}
where the $T_i$ and $V=V^*$ are square matrices of size $L$ which do not depend on any space variable.  Note that $H$ is self-adjoint.

\subsection{Transfer operators}
\label{sec-transferop}

It is intrinsic to the particular form \eqref{eq-genH} that there are two different ways to write $H$ as a Jacobi operators:
\begin{eqnarray*}
H
& = &
(T_1+T_3^*S_2)S_1^*+(T_2^*S_2+T_2S_2^*+V)
+(T_1^*+T_3S_2^*)S_1
\\
& = &
(T_2+T_3S_1){S}_2^*
+(T_1^*S_1+T_1S_1^*+V)
+(T_2^*+T_3^*S_1^*){S}_2
\;.
\end{eqnarray*}
Hence let us denote the coefficient operators by
\begin{equation}
\label{eq-ABdef}
A_1\;=\;T_1+T_3^*S_2\,,\quad
B_1\;=\;T_2^*S_2+T_2S_2^*+V
\,,\quad
A_2\;=\;T_2+T_3S_1\,,\quad
B_2\;=\;T_1^*S_1+T_1S_1^*+V
\,.
\end{equation}
Here we view $A_1$ and $B_1$ as operators on the fiber Hilbert space $\Hh_2=\ell^2(\ZM)\otimes \CM^L$ and $A_2$ and $B_2$ as operators on the other fiber Hilbert space $\Hh_1=\ell^2(\ZM)\otimes \CM^L$ where the $\ZM$ in $\Hh_2$ corresponds to the $2$-direction and that in $\Hh_1$ to the $1$-direction. This corresponds to decompositions of a vector $\psi=(\psi_{n_1,n_2})_{n_1,n_2\in\ZM}\in\Hh$ once into $\psi=(\psi_{n_1})_{n_1\in\ZM}$ with $\psi_{n_1}=(\psi_{n_1,n_2})_{n_2\in\ZM}\in\Hh_2$ and once into $\psi=(\psi_{n_2})_{n_2\in\ZM}$ with $\psi_{n_2}=(\psi_{n_1,n_2})_{n_1\in\ZM}\in\Hh_1$. The total Hilbert space is then given by $\Hh=\ell^2(\ZM)\otimes\Hh_2$ or $\Hh=\ell^2(\ZM)\otimes\Hh_1$ correspondingly. For sake of notational convenience  we will write $j'=j+1\!\!\!\mod\!2$ for $j=1,2$, namely $1'=2$ and $2'=1$. Thus given $j$, $j'$ is not a new index. This allows to state that $A_j$, $B_j$ act on $\Hh_{j'}$, and that
$$
H\;=\;A_j S_j^*+B_j+A_j^* S_j\;,
\qquad
j=1,2
\;.
$$
In order to avoid inessential technical problems, we will suppose that the following holds:

\vspace{.2cm}

\noindent {\bf Hypothesis:} {\it $A_1$ and $A_2$ are invertible} 

\vspace{.2cm}

There are various ways to guarantee this hypothesis. For example, it holds if $T_1$ and $T_2$ are invertible and $T_1^{-1}T_3$ and $T_2^{-1}T_3$ have norm less than $1$. Beneath other things this implies that $H$ has no flat bands (by Proposition~\ref{prop-Ainvertible} in the Appendix) and that the transfer operators are well-defined bounded operators. When $H$ is obtained from an operator on a triangular and hexagonal lattice, the rotational symmetry has to be slightly broken in order to assure that the hypothesis holds (see Section~\ref{sec-examples} for details). If one treats these systems with rotational symmetry, the transfer operators become unbounded due to the divergence of fibers in the direct integral representation discussed below. This can be dealt with, but we choose not to.

\vspace{.2cm}

The two ways to decompose states each allow to study solutions of $H\psi=E\psi$ using the two transfer operators defined by 
\begin{equation}
\label{eq-transferopdef}
\Tt^E_j
\;=\;
\begin{pmatrix}
(E\,{\bf 1}-B_j)A_j^{-1}
& -\,A_j^*
\\
A^{-1}_j & 0
\end{pmatrix}\;.
\end{equation}
We will consider $\Tt^E_j$ as an operator acting on $\Hh_{j'}\oplus\Hh_{j'}$.  For real $E$, these operators are $\Jj$-unitary in the sense of Krein \cite{Kre}, namely they satisfy
\begin{equation}
\label{eq-symplectictransferop}
\Tt^*\Jj\Tt\;=\;\Jj\;,
\qquad
\Jj\;=\;
\begin{pmatrix}
0 & -\one \\ \one & 0
\end{pmatrix}
\;.
\end{equation}
Now suppose that $\psi\in\Hh$ is a solution of the Schr\"odinger equation $H\psi=E\psi$. Using the transfer operator $\Tt_j^E$ and the decomposition of a state $\psi=(\psi_{n})_{n\in\ZM}$ with a fiber vector $\psi_n\in\Hh_{j'}$, one can rewrite the Schr\"odinger equation as 
\begin{equation}
\label{eq-transferrel}
\begin{pmatrix}
A_j\psi_{n+1} \\ \psi_{n}
\end{pmatrix}
\;=\;
\Tt_j^E\,
\begin{pmatrix}
A_j\psi_{n} \\ \psi_{n-1}
\end{pmatrix}\;.
\end{equation}
We will use the equation $H\psi=E\psi$ also for formal solutions $\psi=(\psi_{n_1,n_2})_{n_1,n_2\in\ZM}$ which are not in the Hilbert space. Also for those formal solutions \eqref{eq-transferrel} remains valid.

\vspace{.2cm}

Many of the results of this paper directly transpose to periodic media with periods $p_1$ and $p_2$ larger than $1$, even though we don't write them out in detail. For example, if the potential $V$ in \eqref{eq-genH} is $p_1$-periodic in the $1$-direction, then there are $p_1$ transfer operators $\Tt^E_{2,l}$ defined as in \eqref{eq-transferopdef}, but with $B_2$ depending on $V_l$ with $l=1,\ldots,p_1$. Then one considers the transfer operators over one periodicity cell:
$$
\Tt^E_{2,p_1,1}
\;=\;
\Tt^E_{2,p_1}
\cdots
\Tt^E_{2,1}
\;.
$$
%

\subsection{Fourier transforms}
\label{sec-Fouriertrafo}

The Hamiltonian \eqref{eq-genH} is translation invariant in $\ZM^2$ and can thus be diagonalized by the discrete Fourier transform
$\Uu=\ell^2(\ZM^2)\otimes \CM^L\to L^2(\TM^2,\frac{d{\bf k}}{4\pi^2})\otimes\CM^L$ where $\TM^2=(-\pi,\pi]^2$ is the Brillouin zone furnished with the normalized Riemannian volume measure $\frac{d{\bf k}}{4\pi^2}$. The Fourier transform is defined by
\begin{equation}
\label{eq-Fourier}
(\Uu\psi)({\bf k})\;=\;
\sum_{{\bf n}\in\ZM^2}\psi_{{\bf n}}\e^{\imath{\bf k}\cdot{\bf n}}\;.
\end{equation}
It is unitary with inverse given by
$$
(\Uu^*\phi)_{\bf n}\;=\;
\int_{\TM^2}\frac{d{\bf k}}{4\pi^2}\;\,\phi({\bf k})\,\e^{-\imath{\bf k}\cdot{\bf n}}\;.
$$
Then $\Uu H\Uu^*=\int^\oplus_{\TM^2}\frac{d{\bf k}}{4\pi^2}\; H({\bf k})$ where
$$
H({\bf k})
\;=\;
\sum_{i=1,2,3}(\e^{\imath k_i}T_i^*+
\e^{-\imath k_i}T_i)
+V
\;,
\qquad
k_3=k_2-k_1
\;,
$$
is an $L\times L$ matrix.  The $L$ eigenvalues of $H({\bf k})$ form the so-called Bloch bands of $H$. In the sequel, it will be useful to study also the two partial Fourier transforms $\Uu_j:\Hh_j\to L^2((-\pi,\pi]),\frac{dk_j}{2\pi})\otimes\CM^{L}$ defined by
\begin{equation}
\label{eq-Fourier1}
(\Uu_j{\psi})(k_j)\;=\;
\sum_{n_j\in\ZM}{\psi}_{n_j}\,e^{\imath n_jk_j}\;.
\end{equation}
Clearly they also extend to $\Hh\cong\Hh_j\otimes\ell^2(\ZM)$ as well as $\Hh_j\oplus\Hh_j$ and we use the same symbol $\Uu_j$ for these extensions. Then $\Uu_j {H}\Uu_j^*=\int^\oplus_{(-\pi,\pi]}\frac{dk_j}{2\pi}\; {H}_{j'}(k_j)$ where now ${H}_{j'}(k_j)$ is an operator on $\Hh_{j'}$. One has
\begin{equation}
\label{eq-Hqdef0}
{H}_{j'}(k_j)
\;=\;
A_{j'}(k_j){S}_{j'}^*
+B_{j'}(k_{j})
+
A_{j'}(k_{j})^*{S}_{j'}
\;,
\end{equation}
where we set
$$
A_1(k_2)\;=\;T_1+e^{\imath k_2}T_3^*\;,
\qquad
A_{2}(k_1)\;=\;T_{2}+e^{\imath k_1}T_3
\;,
\qquad
B_{j'}(k_{j})\;=\;e^{\imath k_{j}} T_{j}^*+e^{-\imath k_{j}}T_{j}+V
\;.
$$
The operators ${H}_{j'}(k_j)$ are two-sided Jacobi matrices with $L\times L$-matrix entries. If the invertibility of the operators $A_1$ and $A_2$ holds, then also the matrices $A_1(k_2)$ and $A_2(k_1)$ are invertible. Therefore, by Proposition~\ref{prop-Ainvertible} the spectra of $H_1(k_2)$ and $H_2(k_1)$ are absolutely continuous (no flat bands) and therefore also $H$ has absolutely continuous spectrum.

\vspace{.2cm}

Also the transfer operators are diagonalized by partial Fourier transforms:
\begin{equation}
\label{eq-transferFourier}
\Uu_j\;\Tt_{j'}^E\;\Uu_j^*
\;=\;
\int^\oplus_{(-\pi,\pi]} \frac{dk_j}{2\pi}\;\Tt^E_{j'}(k_j)
\;,
\end{equation}
with $2L\times 2L$ matrices given by
\begin{equation}
\label{eq-transfermatFourier}
\Tt^E_{j'}(k_j)
\;=\;
\begin{pmatrix}
(E\,{\bf 1}-B_{j'}(k_j))A_{j'}(k_j)^{-1}
& -\,A_{j'}(k_j)^*
\\
A_{j'}(k_j)^{-1} & 0
\end{pmatrix}\;.
\end{equation}
Again the matrices $\Tt^E_{j'}(k_j)$ are $\Jj$-unitary so that their determinant lies on the unit circle.  In fact, $\det(\Tt_{j'}^E(k_j))=\overline{\det(A_{j'}(k_j))}/\det(A_{j'}(k_j))$. As it will be of particular importance for the calculation of the edge spectrum below, let us write out $\Tt^E_2(k_1)$ explicitly:
\begin{equation}
\label{eq-transfermat2}
{\Tt}^E_2(k_1)\;=\;
\begin{pmatrix}
(E\,{\bf 1}-(e^{\imath k_1} T_1^*+e^{-\imath k_1}T_1+V))(T_2+e^{\imath k_1}T_3^*)^{-1}
& -\,(T_2+e^{\imath k_1}T_3^*)^*
\\
(T_2+e^{\imath k_1}T_3^*)^{-1} & 0
\end{pmatrix}\;.
\end{equation}

\vspace{.2cm}

\subsection{Chern numbers and spin Chern numbers}
\label{sec-Chern}

Let $P$ be the spectral projection on one or several bands of $H$. Then $P$ is a smooth projection, namely after Fourier transform it fibers into $\Uu P\Uu^*=\int_{\TM^2}^\oplus d{\bf k}\,P({\bf k})$ in $L^2(\TM^2)\otimes\CM^L$ with projections $P({\bf k})$ in $\CM^L$ depending smoothly on $k$. In particular, $N=\mbox{dim}(P({\bf k}))$ is constant. Locally in $\TM^2$ there exists a differentiable orthonormal family $(\psi_l({\bf k}))_{l=1,\ldots,N}$ in $\CM^{L}$ such that
\begin{equation}
\label{eq-Pvec}
P({\bf k})
\;=\;
\sum_{l=1}^{N}\,|\psi_l({\bf k})\rangle\langle \psi_l({\bf k})|
\;.
\end{equation}
The functions ${\bf k}\in\TM^2\mapsto \psi_l({\bf k})$ cannot all be globally smooth if the Chern number $\Ch(P)$ of $P$ does not vanish. Recall that it is an integer number that can be defined by
\begin{equation}
\label{eq-Cherndef}
\Ch(P)
\;=\;
\frac{1}{2\pi\imath}\;
\int_{\TM^2} d{\bf k}\;\Tr(P({\bf k})[\partial_{k_1}P({\bf k}),\partial_{k_2}P({\bf k})])
\;.
\end{equation}
%

\vspace{.2cm}

In presence of a spin degree of freedom, one can further refine the Chern numbers, by passing to so-called spin Chern numbers \cite{SWSH,Pro}. Let us recall that the spin of the particle is denoted by $s\in\NM/2$. Associated to $s$ is an irreducible representation of $\mbox{\rm SU}(2)$ on $\CM^r$ with $r=2s+1$. Let ${\bf s}=(s^x,s^y,s^z)$ be the $r\times r$ matrices representing the 3 components of the angular momentum operators giving a basis of the Lie algebra $\mbox{\rm su}(2)$. We choose the representation such that $s^x$ and $s^z$ are real, and  $s^y$ is purely imaginary. Now let us first suppose that the Hamiltonian $H$ commutes with $s^z$.  This is characterized by
$$
[H,s^z]\;=\;0
\qquad
\Longleftrightarrow
\qquad
[T_i,s^z]\;=\;0\;\;\mbox{ and }\;\;[V,s^z]\;=\;0
\;.
$$
Then let $\Pi_l$ for $l=-s,-s+1,\ldots,s$ denote the spectral projections of $s^z$ on the corresponding eigenvalue. The spectral projection $P$ of $H$ commutes with the projections $\Pi_l$. Setting $H_l=\Pi_l H\Pi_l$, one then has $H=\sum_{l=-s}^sH_l$, namely the Hamiltonian decomposes into a direct sum of Hamiltonians within the spin sector $l$. Let us set $P_l=\Pi_lP=\Pi_lP\Pi_l$, which is also the spectral projection of $H_l$ onto the same spectral set as $P$. Then the spin Chern numbers are defined by 
\begin{equation}
\label{eq-SQHnumbers}
\SCH_l(P)
\;=\;\Ch(P_l)\;,
\qquad
l=-s,\ldots,s
\;,
\end{equation}
By additivity of the Chern numbers one has $\sum_{l=-s}^{s}\SCH_l(P)=\Ch(P)$. Also the situation with a moderate coupling of a term not commuting with $s^z$ (such as the Rashba term, see Section~\ref{sec-examples}) can be dealt with following an idea of Prodan \cite{Pro}. In fact, suppose that the spectrum of the self-adjoint operator $Ps^zP$ on $P\Hh$ consists of $2s+1$ clusters. Then a smooth orthogonal projection $P_l$ is associated to the $l$th cluster of the spectrum, and one can again use \eqref{eq-SQHnumbers} to define the spin Chern numbers. The same procedure can also be applied to systems without periodicity such as disordered systems. All one has to check is the sufficient regularity of the projections $P_l$, but this follows immediately from a contour integration combined with a Combes-Thomas estimate. The physical significance of the spin Chern numbers will be discussed below (see Theorem~\ref{theo-spincurrentquant}).

\section{Half-space Hamiltonians}
\label{sec-halfmodels}

In this section, the symmetry in the directions $1$ and $2$ is broken. For sake of concreteness, let the physical space be the half-space $\ZM\times\NM$ where the $1$-direction is $\ZM$ and the $2$-direction is $\NM$ (with $0$ incluced). The half-plane operators $\widehat{H}$ are obtained by restricting $H$ to the half-space $\ZM\times\NM$. Hence $\widehat{H}$ acts on $\widehat{\Hh}=\ell^2(\ZM\times\NM)\otimes\CM^L$. Here by restriction we mean Dirichlet boundary conditions on the boundary of $\ZM\times\NM$. This means that the unitary $S_2$ is replaced by a partial isometry $\widehat{S}_2$ defined by $(\widehat{S}_2)^*\widehat{S}_2=\one-|0\rangle\langle 0|$ and $\widehat{S}_2(\widehat{S}_2)^*=\one$. If we set $\widehat{S}_1=S_1$ and $\widehat{S}_3=\widehat{S}_1^*\widehat{S}_2$, then
$$
\widehat{H}\;=\;\sum_{i=1,2,3}(T_i^*\widehat{S}_i+T_i\widehat{S}_i^*)\;+\;V
\;.
$$
In principle, other local and translation invariant boundary conditions are possible. This leads to (minor) changes in the intersection theory below, but we expect all the topological statements to be independent of the choice of boundary condition. 

\vspace{.2cm}

Next the transfer matrices for the half-space are simply defined by replacing $S_j$ by $\widehat{S}_j$. We will mainly work with the transfer operator $\Tt^E_2$ in this paper. It is the same for the half-space and full space operator because $\widehat{S}_1=S_1$. 

\vspace{.2cm}

\subsection{Edge spectrum}
\label{sec-edgespec}

The operator $\widehat{H}$ is still translation invariant in the $1$-direction and can hence be partially diagonalized by $\Uu_1:\ell^2(\ZM\times\NM)\otimes \CM^{L}\to L^2((-\pi,\pi]),\frac{dk_1}{2\pi})\otimes \ell^2(\NM)\otimes\CM^{L}$ defined by the same formula \eqref{eq-Fourier1} as above. Then $\Uu_1 \widehat{H}\,\Uu_1^*=\int^\oplus_{(-\pi,\pi]}\frac{dk_1}{2\pi}\; \widehat{H}_2(k_1)$ where now $\widehat{H}_2(k_1)$ is an operator on $\ell^2(\NM)\otimes\CM^{L}$. One has
\begin{equation}
\label{eq-Hqdef}
\widehat{H}_2(k_1)
\;=\;
(T_2+e^{\imath k_1}T_3^*)\widehat{S}_2^*
+(e^{\imath k_1} T_1^*+e^{-\imath k_1}T_1+V)
+(T_2^*+e^{-\imath k_1}T_3)\widehat{S}_2
\;.
\end{equation}
This operator $\widehat{H}_2(k_1)$ is given by the same formula as the periodic Jacobi matrix ${H}_2(k_1)$ with ${S}_2$ replaced by $\widehat{S}_2$. Hence $\widehat{H}_2(k_1)$ is the half-space Jacobi matrix and the general theory to calculate its spectrum as developed in Appendix~\ref{app-periodic} applies.  We recall our hypothesis on the invertibility of  $A_2$ so that $A_2(k_1)=T_2+e^{\imath k_1}T_3$ is also invertible. Then Proposition~\ref{prop-Ainvertible} in Appendix~\ref{app-periodic} implies that $\widehat{H}_2(k_1)$ has no embedded eigenvalues and the bound states lie only in the gaps of $H_2(k_1)$. These bound states of $\widehat{H}_2(k_1)$ constitute the edge spectrum of $\widehat{H}$ defined next.

\begin{defini}
\label{def-edgespectrum}
$E\in\RM $ belongs to the edge spectrum $\sigma^{\mbox{\rm\tiny e}}(\widehat{H})$ if and only if there is a $k_1\in\TM^1$ such that $E$ is an eigenvalue of $\widehat{H}_2(k_1)$.
\end{defini}

Because $\widehat{H}_2(k_1)$ depends analytically on $k_1$, the edge spectrum actually consists of bands $E^{\mbox{\rm\tiny e}}_n(k_1)$ where $n$ is an index labeling the edge bands. We choose the labeling such that $E^{\mbox{\rm\tiny e}}_n(k_1)$ is analytic in $k_1$. A bound state of $\widehat{H}_2(k_1)$ can be at most $L$-fold degenerate, so that at most $L$ of the edge bands can intersect. There can be flat edge bands, that is, an $n$ such that $E^{\mbox{\rm\tiny e}}_n(k_1)$ is independent of $k_1$ \cite{FWNK,NGK}. 

\vspace{.2cm}

Given an energy $E\in\sigma^{\mbox{\rm\tiny e}}(\widehat{H})$, there is a $k_1\in\TM^1$ and a state $\psi_{k_1}\in\widehat{\Hh}_2=\ell^2(\NM)\otimes\CM^L$ such that  $\widehat{H}_2(k_1)\psi_{k_1}=E\psi_{k_1}$.  Then $\psi(n_1,n_2)=e^{\imath k_1 n_1}\psi_{k_1}(n_2)$ is an edge state with quasi-momentum $k_1$ along the boundary. This state $\psi$ is not square-integrable, but falls off from the boundary (that is, it decays in the variable $n_2$). Such edge states are also called Tamm states \cite{Tam} or Shockley states \cite{Sho}.  Let us point out that the edge spectrum $\sigma^{\mbox{\rm\tiny e}}(\widehat{H})$ can have a non-trivial intersection with the Bloch bands $\sigma(H)$.  

\vspace{.2cm}

The edge spectrum is given by the bound states of the half-sided Jacobi matrix $\widehat{H}_2(k_1)$ which can be detected as poles of the Green matrix of $\widehat{H}_2(k_1)$:
$$
\widehat{G}^E(k_1)
\;=\;
\pi_0^*(\widehat{H}_2(k_1)-E)^{-1}\pi_0
\;.
$$
Here $\pi_n:\CM^{L}\to\ell^2(\NM,\CM^{L})$ is the partial isometry on the $n$th site.  Now let us set
\begin{equation}
\label{eq-Uformulas}
U^E(k_1)\;=\;
\left(\widehat{G}^E(k_1) \,+\,\imath\,\one\right)
\left(\widehat{G}^E(k_1) \,-\,\imath\,\one\right)^{-1}
\;=\;
\left(\one \,+\,\imath\,\widehat{G}^E(k_1)^{-1}\right)
\left(\one \,-\,\imath\,\widehat{G}^E(k_1)^{-1}\right)^{-1}
\;.
\end{equation}
The following result is entirely proved in the appendix. It only deals with the spectral analysis of the half-sided Jacobi matrix $\widehat{H}_2(k_1)$ for fixed $k_1$.

\begin{theo}
\label{theo-edgecalc}
Let $E$ be in a gap of the spectrum of $H_2(k_1)$. The $L\times L$ matrix $U^E(k_1)$ is unitary and $\frac{1}{\imath} U^E(k_1)^*\partial_EU^E(k_1)<0$ so that the eigenvalues of $U^E(k_1)$ rotate in the negative sense as a function of $E$. Moreover, 
$$
\mbox{\rm multiplicity of }E\;\mbox{\rm as eigenvalue of }\widehat{H}_2(k_1)
\;=\;
\mbox{\rm multiplicity of }1\;\mbox{\rm as eigenvalue of }U^E(k_1)
\;.
$$
\end{theo}

\vspace{.2cm}

Theorem~\ref{theo-edgecalc} is particularly useful because there is an efficient way to calculate the unitary $U^E(k_1)$ defined in \eqref{eq-Uformulas} which only invokes the transfer matrix $\Tt^E_2(k_1)$ and not the Green matrix. (If the system is $p_1$-periodic in the $1$-direction, then one rather uses $\Tt^E_{2,p_1,1}(k_1)$.) This is discussed in detail in Appendix~\ref{app-periodic}, but now briefly transposed into the present context. In fact, for $E$ not in the spectrum of the two-sided Jacobi matrix $H_2(k_1)$, the transfer matrices $\Tt^E_2(k_1)$ explicitly given in \eqref{eq-transfermat2} are hyperbolic (no eigenvalues of modulus $1$) and therefore the generalized eigenspaces for eigenvalues of modulus strictly less than $1$ constitute an $L$-dimensional $\Jj$-Lagrangian plane in $\CM^{2L}$. Let a basis of this space form the column vectors of a $2L\times L$ matrix $\Phi^E(k_1)$. Then the stereographic projection of this plane is equal to the unitary $U^E(k_1)$, namely:
\begin{equation}
\label{eq-UPhiCalc}
U^E(k_1)\;=\;\binom{1}{\imath}^*\Phi^E(k_1)\left( \binom{1}{-\imath}^*\Phi^E(k_1)\right)^{-1}
\;.
\end{equation}
The equality follows from Weyl-Titchmarch theory which implies that the span of $\Phi^E(k_1)$ coincides with the span of $\binom{\widehat{G}^E(k_1)}{-\one}$. 

\vspace{.2cm}

The edge state invariants defined in the next section are expressed in terms of  the dependence of the unitaries $U^E(k_1)$ in $k_1$, still for $E$ in a gap of $H_2(k_1)$. For that purpose, the following result will be relevant.

\begin{proposi}
\label{prop-deriveeq}
Let $E$ be in a gap of the spectrum of $H$. Suppose that $k_1$ and $n$ are such that the $n^{\mbox{\small th}}$ band of edge states satisfies $E^{\mbox{\rm\tiny e}}_n(k_1)=E$. Let $e^{\imath \theta^E_n(k_1)}=1$ be the corresponding eigenvalue of $U^E(k_1)$ as given by {\rm Theorem~\ref{theo-edgecalc}}, also chosen to be analytic in $k_1$. Then 
$$
\partial_{k_1}\theta^E_n(k_1)
\;=\;
2\,\partial_{k_1}E^{\mbox{\rm\tiny e}}_n(k_1)
\;.
$$
\end{proposi}

\noindent {\bf Proof.}  Let us first assume that the eigenvalue is simple. Hence there is a unit vector $v^E(k_1)\in\CM^{L}$ such that $\widehat{G}^E(k_1)^{-1}v^E(k_1)=0$ which is unique up to a phase factor. This is equivalent to $\widehat{H}_2(k_1)v^E(k_1)=Ev^E(k_1)$ and $U^E(k_1)v^E(k_1)=v^E(k_1)$ ({\it cf.} the second proof of Theorem~\ref{theo-rotprop} in the appendix). Now 
$$
\partial_{k_1}\theta^E_n(k_1)
\;=\;
\langle v^E(k_1)|
\,\tfrac{1}{\imath}\; U^E(k_1)^*\partial_{k_1} U^E(k_1)\,
|v^E(k_1)\rangle
\;.
$$
Deriving the second equation of  \eqref{eq-Uformulas}, one finds
$$
\frac{1}{\imath}\; U^E(k_1)^*\partial_{k_1} U^E(k_1)
\; = \; 
2\,\left(\bigl(\one-\imath\,\widehat{G}^E(k_1)^{-1}\bigr)^{-1}\right)^*
\,\bigl(-\partial_{k_1}\widehat{G}^E(k_1)^{-1}\bigr)
\,\bigl(\one-\imath\,\widehat{G}^E(k_1)^{-1}\bigr)^{-1}
\;.
$$
As $(\one-\imath\,\widehat{G}^E(k_1)^{-1})^{-1}v^E(k_1)=v^E(k_1)$, it follows that
$$
\partial_{k_1}\theta^E_n(k_1)
\;=\;
2\;\langle v^E(k_1)|
\,\bigl(-\partial_{k_1}\widehat{G}^E(k_1)^{-1}\bigr)\,
|v^E(k_1)\rangle
\;=\;
2\;\partial_{k_1}E^{\mbox{\rm\tiny e}}_n(k_1)
\;,
$$
where the last equality holds because the vanishing eigenvalue of $\widehat{G}^E(k_1)^{-1}$ is $E^{\mbox{\rm\tiny e}}_n(k_1)-E$. If the eigenvalue is degenerate, the above calculation is done for each analytic branch.
\hfill $\Box$

\subsection{The edge index and the spin edge indices}
\label{sec-invariant}

\begin{defini}
\label{def-edgeinv}
The edge index $\EI(E)$ at an energy $E$ in the gap of $H$ is given by the winding number of the closed analytic path $k_1\in \SM^1\mapsto U^E(k_1)$:
\begin{equation}
\label{eq-edgeinvariant}
\EI(E)
\;=\;
\int^\pi_{-\pi}\frac{dk_1}{2\pi\imath}\;\partial_{k_1}\,\ln\left(\det(U^E(k_1))\right)
\;.
\end{equation}
\end{defini}

The following result shows that $\EI(E)$ is locally independent of $E$. This implies, in particular, that $\EI(E)=0$ for $E$ below or above the spectrum of $H$ because $U^E(k_1)\to -\one$ as $|E|\to\infty$ uniformly in $k_1$.

\begin{proposi}
\label{prop-derivative}
If $\Delta\subset \RM$ is a gap of the spectrum of $H$, then $E\in\Delta\mapsto \EI(E)$ is a constant and integer-valued. It is equal to the number of weighted crossings of the eigenvalues of $U^E(k_1)$ by $1$, namely 
\begin{equation}
\label{eq-Maslov}
\EI(E)
\;=\;
\sum_{k_1\in\SM^1}
\nu(k_1)
\end{equation}
where the signature $\nu(k_1)$ of $k_1$ is equal to the number of eigenvalues of $U^E(k_1)$ passing by $1$ at $k_1$ in the positive sense minus the number of eigenvalues passing in the negative sense {\rm (}thus there are only a finite number of non-vanishing summands in {\rm \eqref{eq-Maslov})}. Moreover, the formula {\rm \eqref{eq-edgeinvariant}} can be further modified by replacing $E$ by any differentiable path $k_1\in \SM^1\mapsto E(k_1)$ such that $E(k_1)$ remains in the same gap of $H_2(k_1)$.
\end{proposi}

\noindent {\bf Proof.}  Within a gap $U^E(k_1)$ is well-defined for all $k_1$. Therefore all statements follow immediately from general principles on winding numbers.
\hfill $\Box$

\vspace{.2cm}

The index $\EI(E)$ is also equal to the Bott-Maslov index of the closed path $k_1\in \SM^1\mapsto \Phi^E(k_1)$ of $\Jj$-Lagrangian planes. Actually, the r.h.s. of \eqref{eq-Maslov} is one way to define the Bott-Maslov index as an intersection number (see {\it e.g.} \cite{SB} for details). 
%
%
The following rather obvious result connects the edge index to the edge currents along the boundary carried by states within $\Delta$, calculated just as in \cite{SKR,KRS}. Therefore the edge index is equal to the edge Hall conductance as defined in \cite{Hat,KRS}.

\begin{proposi}
\label{prop-edgecurrent}
Let the interval $\Delta\subset \RM$ be in a gap of the spectrum of $H$. Then
$$
\widehat{\Tt}\bigl(\chi_\Delta(\widehat{H})\,\imath[X_1,\widehat{H}]\bigr)
\;=\;
|\Delta|\;
\EI(E)
\;,
\qquad
E\in\Delta
\;,
$$
where $\chi_\Delta$ is the indicator function, $X_1$ the $1$-component of the position operator and $\widehat{\Tt}$ the trace per unit volume in the $1$-direction and usual trace in the $2$-direction, namely for any periodic operator $A$ on $\ell^2(\ZM^2,\CM^L)$ restricted to $\widehat{A}$ on $\ell^2(\ZM\times\NM,\CM^L)$:
$$
\widehat{\Tt}(\widehat{A}\,)
\;=\;
\sum_{n_2\geq 0}\;\Tr\bigl(\langle 0,n_2|\,\widehat{A}\,|0,n_2\rangle\bigr)
\;.
$$
\end{proposi}

\noindent {\bf Proof.}  
Due to the translation invariance, one has
$$
\widehat{\Tt}\bigl(\chi_\Delta(\widehat{H})\,\imath[X_1,\widehat{H}]\bigr)
\;=\;
\sum_n\,
\int^\pi_{-\pi} dk_1\;
\chi_\Delta(E^{\mbox{\rm\tiny e}}_n(k_1))\;\partial_{k_1}E^{\mbox{\rm\tiny e}}_n(k_1)
\;,
$$
where the sum runs over all edge bands in $\Delta$. Now let us use the linearity of this expression to split $\Delta$ into small intervals such that in each of them, say $\Delta'\subset\Delta$, the derivative $\partial_{k_1}E^{\mbox{\rm\tiny e}}_n(k_1)$ only vanishes if $E^{\mbox{\rm\tiny e}}_n(k_1)$ is a boundary point of $\Delta'$. By the fundamental theorem, the contribution of  each edge band is either $|\Delta'|$ or $-|\Delta'|$ pending on the sign of $\partial_{k_1}E^{\mbox{\rm\tiny e}}_n(k_1)$ on the crossing (which is now clearly defined in $\Delta'$). But the sign of $\partial_{k_1}E^{\mbox{\rm\tiny e}}_n(k_1)$ is equal to the sign of $\partial_{k_1}\theta^E_n(k_1)$ at the crossing by Proposition~\ref{prop-deriveeq}, and therefore equal to the signature $\nu(k_1)$ of the crossing (if there are multiple crossings at $k_1$, one has to sum up all contributions). Hence Proposition~\ref{prop-derivative} concludes the proof.
\hfill $\Box$

\vspace{.2cm}

If the Hamiltonian $\widehat{H}$ commutes with $s^z$, one can further refine the edge index similarly as the Chern number of a projection is split into the spin Chern numbers. In fact, if $[\widehat{H},s^z]=0$ the Hamiltonian decomposes into a direct sum $\widehat{H}=\sum_{l=-s}^s\widehat{H}_l$ of Hamiltonians $\widehat{H}_l=\widehat{\Pi}_l\widehat{H}\widehat{\Pi}_l$ acting on the $l$th spin sector $\widehat{\Hh}_l=\widehat{\Pi}_l\widehat{\Hh}$. Then the $l$th spin edge index $\SEI_l(E)$ at energy $E$ is defined as the edge index of the Hamiltonian $\widehat{H}_j$ at energy $E$. Then the same calculation leading to Proposition~\ref{prop-edgecurrent} carries through and therefore one has, for $E\in\Delta$,
\begin{equation}
\label{eq-edgespincurcalc}
\widehat{\Tt}\bigl(\widehat{\Pi}_l\,\chi_\Delta(\widehat{H})\,\imath[X_1,\widehat{H}]\bigr)
\;=\;
|\Delta|\;
\SEI_l(E)
\;.
\end{equation}

\vspace{.2cm}

\subsection{Link between edge indices and Chern numbers}
\label{sec-connection}

The following result generalizes the result of Hatsugai for the Harper model \cite{Hat}, for which transfer matrices are only of size $2$, namely $L=1$. It follows directly from the techniques of \cite{KRS}, which we don't repeat here in detail.  We expect that a more direct proof using only Bloch theory can be obtained.
Another proof can likely be obtained from the approach in  \cite{EG}.

\begin{theo}
\label{theo-edgechannelcount}
Let $E_-<E_+$ be two energies lying in two gaps of $H$ and let $P$ be the spectral projection of $H$ on $[E_-,E_+]$. Then
$$
\Ch(P)\;=\;\EI(E_+)\,-\,\EI(E_-)
\;.
$$
If, moreover, $[H,s^z]=0$, then for $l=-s,\ldots,s$, one has
$$
\SCH_l(P)\;=\;\SEI_l(E_+)\,-\,\SEI_l(E_-)
\;.
$$
\end{theo}

\noindent {\bf Proof.} For $L=1$ this already explicitly contained in \cite{KRS} which also covers periodic systems. For larger $L$ one has to tensor the algebras in \cite{KRS} by the $L\times L$ matrices. As the arguments in \cite{KRS} are only of $K$-theoretic nature and are thus stable w.r.t. tensorizing with compact operators, also the case of larger $L>1$ is covered.
\hfill $\Box$

\subsection{Spin edge currents and spin Chern numbers}
\label{sec-spincurrent}

For a Hamiltonian $H$ commuting with $s^z$ one can refine Theorem~\ref{theo-edgechannelcount}. For these operators the spin Chern numbers are well-defined and have a clear physical significance due the following result, notably they determine the quantization numbers of the spin edge currents associated to the spin current operator $s^z \dot{X}_1=\imath[s^zX_1,\widehat{H}]$.

\begin{theo}
\label{theo-spincurrentquant}
Let $H$ be a Hamiltonian of the form {\rm \eqref{eq-genH}} which commutes with $s^z$. Let the interval $\Delta\subset \RM$ be in a gap of the spectrum of $H$. Then for each $l\in\{-s,\ldots,s\}$ the spin edge currents with spin $l$ are quantized and given by the spin Chern numbers according to the identity:
$$
\widehat{\Tt}\bigl(\widehat{\Pi}_l\,\chi_\Delta(\widehat{H})\,\imath[s^zX_1,\widehat{H}]\bigr)
\;=\;
l\;|\Delta|\;\SCH_l(P)
\;,
$$
where $P$ is the spectral projection of $H$ on all bands below $\Delta$, $\widehat{\Pi}_l=\chi_{\{l\}}(s^z)$ is the spectral projection of $s^z$ onto the eigenvalue $l$ and  $\widehat{\Tt}$ is defined in {\rm Proposition~\ref{prop-edgecurrent}}.
\end{theo}

\noindent {\bf Proof.} As $[H,z^z]=0$, the Hamiltonian can be decomposed into a direct sum $H=\sum_{l=-s}^s H_l$ where $H_l=\Pi_lH\Pi_l$. For each operator $H_l$, one first applies Proposition~\ref{prop-edgecurrent} and then Theorem~\ref{theo-edgechannelcount}, or more precisely, summing the identity of Theorem~\ref{theo-edgechannelcount} over all bands below $\Delta$ and using that edge index always vanishes below the spectrum.
\hfill $\Box$

\section{Invariants for time reversal invariant models}
\label{sec-TRI}

\subsection{The time reversal operator}
\label{sec-TRIop}

The time reversal operator $\Theta$  on $\ell^2(\ZM^2,\CM^L)$  is given by the complex conjugation $K$ and a rotation of each spin component by 180 degrees. If $K$ also acts on the spin degree of freedom so that it contains the rotation of $s^x$ and $s^z$, one thus has
$$
(\Theta \psi)_n\;=\;I\;\overline{\psi_n}\;,
\qquad
I\;=\;\e^{\imath \pi s^y}
\;.
$$
Hence $\Theta$ is an anti-unitary operator, namely $\Theta$ is anti-linear and satisfies $\Theta^*\Theta=\Theta\Theta^*=\one$ where $\Theta^*$ is defined by $\langle \psi|\Theta^*\phi\rangle=\langle\phi| \Theta\,\psi\rangle$. Clearly $\Theta^*=\Theta^{-1}$. Moreover, $I$ is real and one has $\Theta^2=e^{2\pi\imath s^y}K^2=(-1)^{2s}$ and $I^*=(-1)^{2s}I$ because the spectrum of $s^y$ is $\{-s,-s+1,\ldots,s\}$.  One calls the time-reversal operator and the spin even or odd depending whenever $\Theta^2=1$ and $\Theta^2=-1$ respectively. Another relation of use below is that the spin transforms like angular momentum:
\begin{equation}
\label{eq-spinrev}
\Theta^{-1}\,{\bf s}\,\Theta\;=\;-\,{\bf s}\;.
\end{equation}

\vspace{.2cm}

The time reversal operator $\Theta$ now also acts fiberwise on the spin degree of freedom $\CM^r$ of the Hilbert spaces $\Hh=\ell^2(\ZM^2)\otimes \CM^R\otimes\CM^r$ and $\widehat{\Hh}=\ell^2(\ZM\times \NM)\otimes \CM^R\otimes\CM^r$. It also acts naturally on tensor products. All these extensions of $\Theta$ will also be denoted by $\Theta$, and they are all strictly local in the space variable $\ZM^2$ or $\ZM\times\NM$. 

\vspace{.2cm}

The Hamiltonian $H$ is called time reversal invariant (TRI) if
$$
\Theta^{-1}\,H\,\Theta\;=\;H\;.
$$
Using complex conjugation this can also be rewritten as $I^*\overline{H}I=H$. As $H$ is of the form \eqref{eq-genH} and the shift operators $S_i$ are real and independent of spin, TRI is equivalent to having
$$
I^{-1}\,\overline{T_i}\,I\;=\;T_i\;,
\qquad
I^{-1}\,\overline{V}\,I\;=\;V\;.
$$
The TRI is inherited by the half-space operators $\widehat{H}$. We will only consider TRI operators in this section and will exhibit numerous examples of TRI operators in Section~\ref{sec-examples}.

\vspace{.2cm}

Next let us consider the transfer operators $\Tt^E_j$ for real energies $E$ as defined in \eqref{eq-transferopdef}. Because its operator entries are TRI, they also satisfy a TRI relation:
\begin{equation}
\label{eq-transferopTRI}
I^{-1}\,\overline{\Tt^E_j}\;I
\;=\;
\Tt^E_j
\;,
\end{equation}
and similarly for $\widehat{\Tt}^E_j$. Together with \eqref{eq-symplectictransferop}  the transfer operators hence satisfy two relations which, if they were matrices, would imply that the transfer operators are in groups isomorphic to the real symplectic group for even spin and in the classical group $\mbox{SO}^*$ for odd spin. In these groups the spectrum is formed by quadruples or couples on the unit circle, a property inherited by the transfer operators.

\vspace{.2cm}

Under the Fourier transform $\Uu$ and the partial Fourier transform $\Uu_j$, the time reversal symmetry is implemented as follows:
$$
(\Uu\,\Theta\,\Uu^*\psi)({\bf k})\;=\;I\,\overline{\psi(-{\bf k})}\;,\qquad
(\Uu_j\,\Theta\,\Uu_j^*{\psi})(k_j)\;=\;I\,\overline{{\psi}(-k_j)}\;.
$$
Hence the TRI of $H$ leads to
$$
I^*\,\overline{H(-{\bf k})}\,I\;=\;H({\bf k})\;,
\qquad
I^*\,\overline{{H}_{j'}(-k_j)}\,I\;=\;{H}_{j'}(k_j)\;.
\qquad
I^*\,\overline{\widehat{H}_{j'}(-k_j)}\,I\;=\;\widehat{H}_{j'}(k_j)\;.
$$
Thus only at the so-called TRI points with $k_j\in\{0,\pi\}$, the matrices $H({\bf k})$, ${H}_{j'}(k_j)$ and $\widehat{H}_{j'}(k_j)$ are TRI. Similarly, for the transfer matrices one has
\begin{equation}
\label{eq-transferTRI}
I^*\,\overline{\Tt^E_{j'}(-k_j)}\,I\;=\;\Tt_{j'}^{E}(k_j)\;.
\end{equation}
Again these are only TRI $\Jj$-unitaries if $k_j\in\{0,\pi\}$. 

\begin{proposi}
The multiplicity of the edge spectrum of an odd {\rm TRI} Hamiltonian is even.
\end{proposi}

\noindent {\bf Proof.} The reason is that the edge spectrum is defined via the point spectrum of $\widehat{H}_2(k_1)$. But if at some energy there is an eigenvector of $\widehat{H}_2(k_1)$, then there is also an eigenvector with same energy for $\widehat{H}_2(-k_1)$.
\hfill $\Box$

\vspace{.2cm}

In the following, it will be important to consider a grading of the Hilbert space $\Hh$ which is adapted to the TRI operator:
\begin{equation}
\label{eq-splitting}
\Hh\;=\;\Hh_+\oplus\Hh_-\;,
\qquad
\Theta\,\Hh_\pm\;=\;\Hh_\mp\;.
\end{equation}
Then $\Hh_+$ and $\Hh_-$ are identified with $\Hh_+\oplus 0$ and $0\oplus\Hh_-$ and the projections on these subspaces will be denoted by $\Pi_+$ and $\Pi_-$. Moreover, it will be required that the Hamiltonian is diagonal in this grading, namely $H=H_++H_-$ wtih $H_\pm=\Pi_\pm H\Pi_\pm$.  Similarly one can proceed with $\widehat{\Hh}=\widehat{\Hh}_+\oplus\widehat{\Hh}_-$ with corresponding projections $\widehat{\Pi}_+$ and $\widehat{\Pi}_-$.  There may be no such splitting ({\it e.g.} in the case of even TRI systems) or many of them (for odd TRI systems), as will become apparent in the following example.  Let us suppose that $[H,s^z]=0$. By \eqref{eq-spinrev}, one has $\Theta^{-1}\Pi_l\Theta=\Pi_{-l}$. Therefore, by choosing from each pair $\Pi_l,\Pi_{-l}$ one to be a summand of $\Pi_+$, a TRI splitting is obtained if $s$ is odd (odd TRI) because then there is no $\Pi_0$ which cannot be further split. For even TRI exactly this happens and thus there is no splitting. For odd TRI, one may  therefore set
$$
\Pi_\pm\;=\;\sum_{\pm l>0}\;\Pi_l
\;,
$$
but actually this is just one of $2^r$ possible choices. All the above can be done in the same manner for $\widehat{\Hh}$. 


\subsection{The edge $\ZM_2$-invariant}
\label{sec-edgeZ2inv}

Let us first show explicitly that the edge index $\EI(E)$ vanishes for every TRI Hamiltonian and energy $E$ in one of its gaps. Indeed, from $\widehat{H}_{2}(-k_1)=I^*\overline{\widehat{H}_2(k_1)}I$ follows $\widehat{G}^E(-k_1)=I^*\overline{\widehat{G}^E(k_1)}I=I^*\widehat{G}^E(k_1)^tI$ so that by \eqref{eq-Uformulas}
\begin{equation}
\label{eq-UTRI}
U^E(-k_1)\;=\;I^*\,U^E(k_1)^t\,I
\;.
\end{equation}
Thus, because $I$ is unitary and the determinant invariant under transposition,
$$
\partial_{k_1}\ln\left(\det(U^E(-k_1))\right)
\;=\;
-\,\partial_{k_1}\ln\left(\det(U^E(k_1))\right)
\;,
$$
which implies that $\EI(E)=0$. On the other hand, \eqref{eq-UTRI} also implies that the spectra of $U^E(k_1)$ and $U^E(-k_1)$ coincide. This is trivial for $k_1=0$ and $k_1=\pi$, but in the case of odd TRI Kramers degeneracy there is another crucial consequence of \eqref{eq-UTRI} at these TRI points: the multiplicity of the spectra of $U^E(0)$ and $U^E(\pi)$ are even. In fact, from $I^*U^tI=U$ and $Uv=e^{\imath\varphi}v$ follows $U^tIv=e^{\imath\varphi}v$ so that $I^*UI\overline{v}=e^{\imath\varphi}I\overline{v}$. Moreover, $v$ and $I\overline{v}$ are linearly independent because
$$
\langle v|I\overline{v}\rangle
\;=\;
\overline{\langle I\overline{v}|v\rangle}
\;=\;
\langle Iv|\overline{v}\rangle
\;=\;
-\,\langle v|I\overline{v}\rangle
\;=\;
0\;.
$$
This degeneracy makes the following definition possible.

\begin{defini}
\label{def-edgeZ2inv}
The edge $\ZM_2$-index $\EI_2(E)\in\ZM_2$ at an energy $E$ in the gap of a Hamiltonian $H$ with odd {\rm TRI} is equal to the number of crossings of the eigenvalues of $k_1\in[0,\pi]\mapsto U^E(k_1)$ with $1$ counted with their multiplicity for $k_1\in(0,\pi)$ and half their multiplicity at $k_1=0$ and $k_1=\pi$, all calculated modulo $2$.
\end{defini}

Let us point out that though crossings at $k_1=0$ and $k_1=\pi$ are only counted with half their multiplicity, the contribution of each of these two points is integer due to the Kramers degeneracy mentioned above. Therefore the total number of crossings as counted in Definition~\ref{def-edgeZ2inv}  is indeed integer valued. 

\begin{proposi}
\label{prop-edgeZ2inv} Let $H$ be a Hamiltonian with odd {\rm TRI} and $E$ in a gap of $H$. The edge $\ZM_2$-index $\EI_2(E)$ is a well-defined homotopy invariant and independent of the choice of $E$ in the gap.
\end{proposi}

\begin{figure}
\begin{center}
\includegraphics[height=4cm]{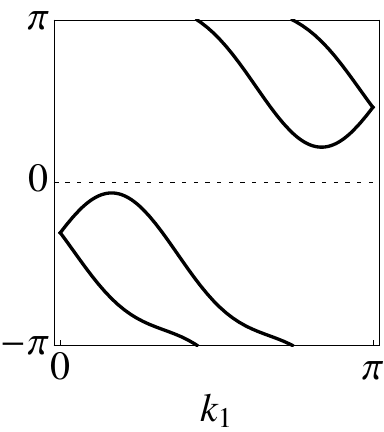}
\hspace{.3cm}
\includegraphics[height=4cm]{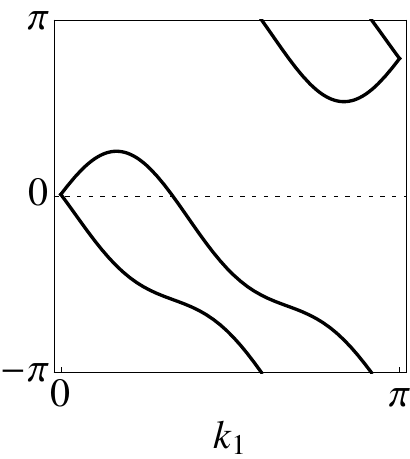}
\hspace{.3cm}
\includegraphics[height=4cm]{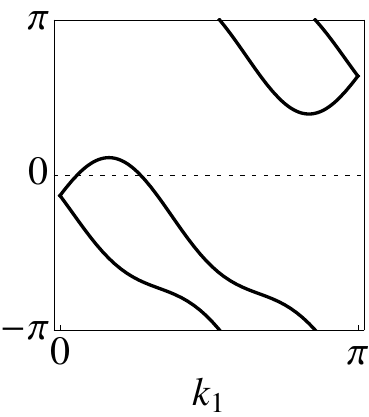}
\caption{\sl Schematic representation of homotopies of one double branch (which is simply given by a shift). One readily sees that the number of intersections counted as in {\rm Definition~\ref{def-edgeZ2inv}} changes, but not modulo $2$.}
\label{fig-Z2def}
\end{center}
\end{figure}

\noindent {\bf Proof.}  
Let $\Xi$ be the set of families of continuous curves $k_1\in[0,\pi]\mapsto e^{\imath\theta_l(k_1)}\in\SM^1$ of phases  with $l=1,\ldots,L$ and $L$ even satisfying at the boundary condition that for $k_1=0$ and $k_1=\pi$ each phase has even multiplicity. Due to the discussion above, the spectrum of $k_1\in[0,\pi]\mapsto U^E(k_1)$ provides one point in $\Xi$.  Now a moment of thought shows that counting the intersections with $1$ modulo $2$ is a homotopy invariant, namely $\Xi$ has exactly two components (see Figure~\ref{fig-Z2def} for an illustration). Furthermore, one can replace the constant curve $1$ by any given continuous curve of vanishing winding number and then calculate the intersection with this new curve without changing the $\ZM_2$-index (actually, just multiply the curves of $\Xi$ by the inverse of the new curve to reduce to the above). Now changing the energy and/or the Hamiltonian such that $E$ remains in a gap does not change the $\ZM_2$-index.
\hfill $\Box$

\begin{proposi}
\label{prop-edgeZ2invMas} Let $H$ be a Hamiltonian with odd {\rm TRI} and $E$ in a gap of $H$. The edge $\ZM_2$-index $\EI_2(E)$ is equal to the Bott-Maslov index of the path $k_1\in [0,\pi]\mapsto \Phi^E(k_1)$ of $\Jj$-Lagrangian planes calculated modulo $2$:
\begin{equation}
\label{eq-Z2Maslov}
\EI_2(E)
\;=\;
\left(\sum_{k_1\in(0,\pi)}
\nu(k_1)
\;+\;
\frac{1}{2}\,\sum_{k_1\in\{0,\pi\}}
\nu(k_1)
\right)\!\!\mod 2
\;,
\end{equation}
where the signature of each intersection is defined as in {\rm Proposition~\ref{prop-derivative}} and thus the sum has only a finite number of non-vanishing summands.
\end{proposi}

\noindent {\bf Proof.} First let point out that weighing crossings at the boundary by a factor $\frac{1}{2}$ is indeed a good convention in the definition of the Bott-Maslov index of non-closed paths because then a natural concatenation holds (see {\it e.g.}  \cite{SB}). Now the only difference between Definition~\ref{def-edgeZ2inv} and \eqref{eq-Z2Maslov} is that crossings are counted with a weight $-1$ or $1$ in the latter. But when calculating only modulo $2$ this does not make any difference.
\hfill $\Box$

\vspace{.2cm}

\begin{figure}
\begin{center}
\includegraphics[height=4cm]{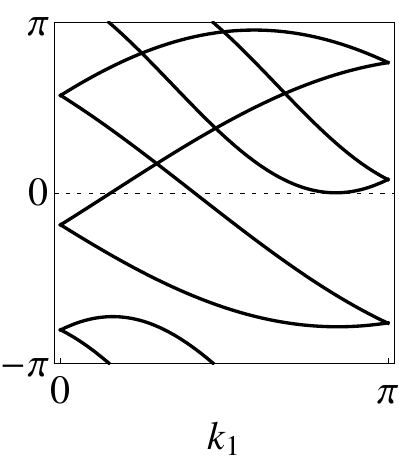}
\hspace{.3cm}
\includegraphics[height=4cm]{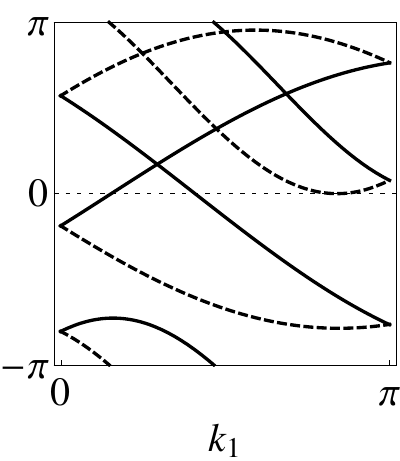}
\hspace{.3cm}
\includegraphics[height=4cm]{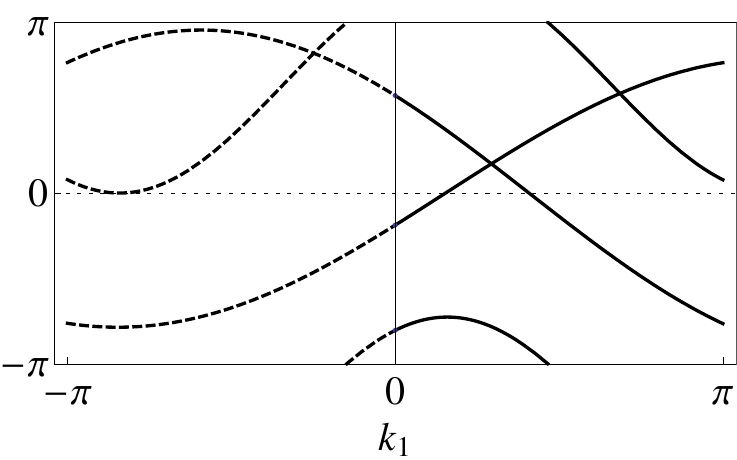}
\caption{\sl Schematic representation of the winding number calculation for the $\ZM_2$-edge index. The left figure shows the edge bands as calculated numerically, or just one point in the set $\Xi$ described in the proof of Proposition~\ref{prop-edgeZ2inv}. On the middle figure a selection of even/odd (non-dashed/dashed) bands has been made. In the right figure the dashed bands are reflected and the winding number modulo $2$ of this figure is the $\ZM_2$-invariant (here it is $0$).}
\label{fig-resolve}
\end{center}
\end{figure}

The $\ZM_2$-index of a point in $\Xi$ can also be calculated as a winding number via the following procedure. Roughly stated, one chooses half of the curves in such a way that they can be concatenated with the other half to closed paths (see Figure~\ref{fig-resolve}). For simplicity let us consider a generic point of $\Xi$ at which each phase at $0$ and $\pi$ is twice degenerate. Start with one of the phases $e^{\imath \theta_1(0)}=e^{\imath \theta_2(0)}$  at $k_1=0$. This gives us phases $e^{\imath \theta_1(\pi)}$ and $e^{\imath \theta_2(\pi)}$  at $k_1=\pi$.  Choose one of the corresponding paths, say $e^{\imath \theta_1(k_1)}$. Then consider the next phase $e^{\imath \theta_3(0)}=e^{\imath \theta_4(0)}$ at  $k_1=0$.  Associated are again two paths $e^{\imath \theta_3(k_1)}$ and $e^{\imath \theta_4(k_1)}$. Choose one, say  $e^{\imath \theta_3(k_1)}$, for which $e^{\imath \theta_3(\pi)}$ is not equal to the previously chosen $e^{\imath \theta_1(\pi)}$. Then iterate choosing while always assuring that no phase at $k_1=\pi$ appears twice beneath the chosen odd ones. Now both the odd and the even paths connect all twice degenerate phases at $0$ and $\pi$. Next $\frac{L}{2}$ paths  $k_1\in(-\pi,\pi]\cong \SM^1\mapsto e^{\imath\theta_{2l}(k_1)}$, $l=1,\ldots,\frac{L}{2}$, are obtained by setting $e^{\imath\theta_{2l}(-k_1)}=e^{\imath\theta_{2l-1}(k_1)}$. Note that these paths are not necessarily closed separately, but if all are considered together one obtains closed paths. Furthermore, the number of crossings with $1$ of the $\frac{L}{2}$ extended paths is clearly equal to the number of crossings of the initial paths. Now, when calculating modulo $2$ it is irrelevant whether one calculates weighted crossings (as in the winding number) or all crossings because the difference is $2$ for each crossing. Thus the sum of the winding numbers of the $\frac{L}{2}$ paths $k_1\in(-\pi,\pi]\cong \SM^1\mapsto e^{\imath\theta_{2l}(k_1)}$ calculated modulo $2$ is equal to $\ZM_2$-index. This also provides an alternative manner to show that the  $\ZM_2$-index is a homotopy invariant.

\vspace{.2cm}

Let us show how a selection of the even and odd curves in the prior remark is naturally given in a situation where $[\widehat{H},s^z]=0$ so that there is a TRI splitting \eqref{eq-splitting} adapted to $\widehat{H}$. This leads to a splitting of the transfer operator ${\Tt}_2^E={\Tt}_{2,+}^E\oplus{\Tt}_{2,-}^E$ and therefore similarly for each of its fibers ${\Tt}_2^E(k_1)$. Moreover, one has
$$
\overline{{\Tt}_{2,-}^E(-k_1)}
\;=\;
{{\Tt}_{2,+}^E(k_1)}
\;,
$$
which is compatible with \eqref{eq-transferTRI}. It in turn implies yet another splitting, compatible with \eqref{eq-UTRI}, namely that 
$$
U^E(k_1)\;=\;U^E_+(k_1)\oplus U^E_-(k_1)
\;,
\qquad
U^E_-(-k_1)^t\;=\;U^E_+(k_1)
\;.
$$
Both $U^E_+(k_1)$ and $U^E_-(k_1)$ are analytic in $k_1$ and are actually associated to the Hamiltonian $\widehat{H}_+$ and $\widehat{H}_-$ respectively and thus allow to calculate their edge states by Theorem~\ref{theo-edgecalc}. Now their spectra have the reflection property $\sigma(U^E_+(k_1))=\sigma(U^E_-(-k_1))$. Hence one may choose the eigenvalues of $U^E_+(k_1)$ for $k_1\in[0,\pi]$ to form the even curves in the terminology of the above, and then the reflected odd curves complete them with $U^E_+(k_1)$ for $k_1\in[-\pi,0]$. Therefore, the $\ZM_2$-index for a Hamiltonian commuting with $s^z$ is given by
\begin{equation}
\label{eq-edgeZ2Maslovcont}
\EI_2(E)
\;=\;
\int^\pi_{-\pi}\frac{dk_1}{2\pi\imath}\;\partial_{k_1}\,\ln\left(\det(U^E_\pm(k_1))\right)
\!\!\mod 2
\;.
\end{equation}
Let us point out that the above argument also implies that this is independent of the choice of the TRI splitting, namely which of the spin eigenvalue pairs $-l,l$ enters into $\widehat{\Hh}_+$ for every $l>0$. 


\subsection{Chern $\ZM_2$-invariant}
\label{sec-SpinChern}

Still let us suppose that $H$ has odd TRI. Then also the spectral projection $P$ on one of its bands is so, namely $\Theta^{-1}P\Theta=P$. Hence we are lead to study TRI fibre bundles on $\TM^2$ satisfying
\begin{equation}
\label{eq-TRIofP}
I^*\overline{P(-{\bf k})}I\;=\;P({\bf k})
\;.
\end{equation}
Using this for the particular case ${\bf k}=0$ shows that $\mbox{dim}(P(0))$ is even if $\Theta$ is odd (by the usual Kramers degeneracy argument applied to $H(0)$) and therefore all $\mbox{dim}(P({\bf k}))$ are even (and equal). Furthermore, the TRI \eqref{eq-TRIofP} of $P$ (even or odd) shows, when replaced into \eqref{eq-Cherndef},
$$
\Ch(P)\;=\;-\,\overline{\Ch(P)}\;=\;0\;.
$$
Hence the Chern number itself is not an interesting invariant for systems with TRI.

\vspace{.2cm}

However, suppose that one has a TRI splitting \eqref{eq-splitting} in which the Hamiltonian is diagonal. Then define $P_\pm=\Pi_\pm P\Pi_\pm$ so that
\begin{equation}
\label{eq-decomp}
P\;=\;
P_+\oplus P_-\;,
\qquad
\Theta^{-1}P_\pm\Theta\;=\;P_\mp
\;.
\end{equation}
If both $P_\pm$ are smooth, their Chern numbers are well-defined. As explained in Section~\ref{sec-TRIop}, one always has a splitting with smooth $P_\pm$ if the commutator $[H,s^z]$ is small. By \eqref{eq-decomp} and the additivity of the Chern number then follows
$$
0
\;=\;
\Ch(P)\;=\;\Ch(P_+)+\Ch(P_-)
\;,
$$
But neither $\Ch(P_+)$ nor $\Ch(P_-)$ is an invariant itself, as it depends on the choice of the splitting and there are many of those ($2^r$ of them if $H$ almost commutes with $s^z$). Nevertheless, similarly as for the edge invariants, changing the splitting only changes $\Ch(P_+)$ by an even number. Therefore the Chern $\ZM_2$-invariant can be defined as
$$
\Ch_2(P)\;=\;\Ch(P_\pm)\!\!\mod 2\;.
$$
We expect this $\ZM_2$-number to be equal the invariant of Kane and Mele \cite{KM}.


\subsection{Edge and Chern $\ZM_2$-indices and spin currents}
\label{sec-edgeZ2invconnect}

This section is about a connection between Chern $\ZM_2$-index and edge $\ZM_2$-invariants which is very much in the spirit of Theorem~\ref{theo-edgechannelcount}. Actually, it is a direct consequence of that theorem in the case where the Hamiltonian commutes with $s^z$. To deal with more general Hamiltonians, we apply a homotopy argument. A concrete homotopy $H(\lambda)$ from a given Hamiltonian $H=H(0)$ to a Hamiltonian $H(1)$ commuting with $s^z$ is given by
\begin{equation}
\label{eq-homotopy}
H(\lambda)\;=\;H\;+\;\frac{\lambda}{2}\,(s^z\,H\,s^z-H)
\;=\;H\;+\;\frac{\lambda}{2}\,[s^z,H]\,s^z
\;,
\qquad
\lambda\in[0,1]
\;.
\end{equation}
Our main hypothesis on $H$ is that its gaps remain open under this homotopy. This is obviously guaranteed if the commutator $[s^z,H]$ is small in operator norm. We expect the following result also to hold without this assumption, but this would require a different proof.

\begin{theo}
\label{theo-edgechannelcount2}
Let $H$ be a Hamiltonian of the form {\rm \eqref{eq-genH}} which has odd {\rm TRI}. 
Let $E_-<E_+$ be two energies lying in two gaps of $H$ which remain open under the homotopy
{\rm \eqref{eq-homotopy}}. Let $P$ be the spectral projection of $H$ on $[E_-,E_+]$. Then
$$
\Ch_2(P)\;=\;\Bigl(\EI_2(E_+)\,+\,\EI_2(E_-)\Bigr)
\!\!\mod 2
\;.
$$
\end{theo}

\noindent {\bf Proof.} By homotopy invariance of $\Ch_2(P)$ and the edge $\ZM_2$-indices it is sufficient to prove the theorem for a Hamiltonian $H$ commuting with $s^z$. For this Hamiltonian one has the TRI splitting \eqref{eq-splitting} in which $H=H_+\oplus H_-$. For $H_+$, Theorem~\ref{theo-edgechannelcount} implies $\Ch(P_+)=\EI(E_+)-\EI(E_-)$ where $\EI(E_\pm)$ are the edge indices associated to $H_+$. But $\Ch_2(P)=\Ch(P_+)\,\mbox{mod} \,2$ by definition, and $\EI_2(E_\pm)=\EI(E_\pm)\,\mbox{mod}\, 2$ by \eqref{eq-edgeZ2Maslovcont}.
\hfill $\Box$

\vspace{.2cm}

The following shows that a non-trivial $\ZM_2$-invariant implies that there are non-vanishing spin edge currents, even though these currents are not quantized any more as for Hamiltonians commuting with $s^z$.

\begin{theo}
\label{theo-spinedgecur}
Suppose that the hypothesis of {\rm Theorem~\ref{theo-edgechannelcount2}} hold and that $\Ch_2(P)=1$. Then for some $l$ and either $E_+$ or $E_-$, the spin edge current of spin $l$ in an interval $\Delta$ in the gap of $E_+$ or $E_-$
$$
\widehat{\Tt}\bigl(\widehat{\Pi}_l\,\chi_\Delta(\widehat{H})\,\imath[s^zX_1,\widehat{H}]\bigr)
$$
is non-vanishing as long as the commutator  $[s^z,H]$ is sufficiently small.
\end{theo}

\noindent {\bf Proof.} If $\Ch_2(P)=1$, then by homotopy the Hamiltonian $H(1)$ which commutes with $s^z$ has a non-vanishing Chern number $\Ch(P_+)$. This implies that for some $l$ the spin Chern number $\SCH_l(P)$  is non-vanishing.  By Theorem~\ref{theo-spincurrentquant} one has $\SCH_l(P)=\SEI_l(E_+)-\SEI_l(E_-)$. Hence either $\SEI_l(E_+)$ or $\SEI(E_-)$ is non-vanishing (still for the operator $H(1)$). Therefore the spin edge current of spin $l$ is non-vanishing by \eqref{eq-edgespincurcalc}. Finally one deforms the Hamiltonian back. As the spin edge current is continuous in the Hamiltonian, it cannot vanish if the path is not too long, namely the commutator $[H,s^z]$ is not too large. 
\hfill $\Box$

\section{Models with spin-orbit interaction on various lattices}
\label{sec-examples}

This section presents the standard two-dimensional tight-binding models, with particular focus on spin-orbit interactions. We restrict to the $1$-periodic case, but it is possible and straightforward to add periodic or disordered potentials. The transfer matrices are written out explicitly and the edge invariants are calculated. 

\subsection{Laplacian on a square lattice}
\label{frepart2dim}

Let $H_0:l^2(\ZM^2)\otimes\CM^r\to l^2(\ZM^2)\otimes \CM^r$ be the discrete Laplacian on a square lattice, namely $H_0=t\,(S_1+S_1^*+S_2+S_2^*)\otimes{\bf 1}$ where $t$ is a real parameter. Thus $R=1$ and
$$
T_1=T_2=t\;,\qquad T_3=V=0\;,
$$
so that
$$
\Tt_1^E(k_2)
\;=\;
\begin{pmatrix}
(E-2t\cos(k_2))t^{-1} & -t \\ t^{-1} & 0
\end{pmatrix}
\;,
\qquad
\Tt_2^E(k_1)
\;=\;
\begin{pmatrix}
(E-2t\cos(k_1))t^{-1} & -t \\ t^{-1} & 0
\end{pmatrix}
\;.
$$
As $H_0$ does not depend on the spin, one may work with spin $0$ and set $r=1$. In the present situation, the unitary $U^E(k_1)$ is just a number on the unit circle. Let us calculate it explicitly. The eigenvalues of the transfer matrix $\Tt^E_2(k_1)$ are
$$
\lambda^E_\pm(k_1)\;=\;
\frac{E-2t\cos(k_1)}{2t}\;\pm\;
\frac{1}{2t}\;\sqrt{(E-2t\cos(k_1))^2-4t^2}
\;,
$$
and the corresponding eigenvectors are
$$
\begin{pmatrix}
t\lambda^E_\pm(k_1) \\ 1
\end{pmatrix}
\;.
$$
For $E>4t$, the smallest (real) eigenvalue is $\lambda^E_-(k_1)$ so that
$$
U^E(k_1)
\;=\;
(t\lambda^E_-(k_1)-\imath)(t\lambda^E_-(k_1)+\imath)^{-1}
\;.
$$
This phase is different from $1$ for all $k_1$ and $E$ outside of the spectrum of $H_0$, so that there are no edge states. Changing the boundary condition from Dirichlet to something else (such that, say, intersections with the phase $e^{\imath\alpha}$), it is, however, possible to produce edge states. Also let us note that  $H_0({\bf k})=2t\,(\cos (k_1)+\cos (k_2))$ and $\widehat{H}_{0,2}(k_1)=t\,(2\cos(k_1)+\widehat{S}_2+\widehat{S}_2^*)$. Hence $\widehat{H}_{0,2}(k_1)$ is given by a sum of a $k_1$-dependent constant term and the discrete Laplacian on the discrete half-line $\NM$. 

\subsection{Spin-orbit Laplacian on a square lattice}

Next let us add to $H_0$ a spin orbit interaction and suppose for sake of concreteness that $r=2$ so that the spin is $\frac{1}{2}$. There are various possible ways to do that (compare \cite{And}), one is given by:
$$
H_\text{\sc so}\;=\;
2\,\imath\,\lambda_\text{\sc so}\,(S_1-S_1^*)
\otimes s^x
+2\,\imath\,\lambda_\text{\sc so}\,(S_2-S_2^*)\otimes s^y
\;,
$$
where ${\bf s}=(s^x,s^y,s^z)$ denote $s^x=\frac{1}{2}\left(\begin{smallmatrix} 0 & 1 \\ 1 & 0 \end{smallmatrix}\right)$,
$s^y=\frac{1}{2}\left(\begin{smallmatrix} 0 & -\imath \\ \imath & 0 \end{smallmatrix}\right)$ and $s^z=\frac{1}{2}\left(\begin{smallmatrix} 1 & 0 \\ 0 & -1 \end{smallmatrix}\right)$ and $\lambda_\text{\sc so}$ is a real coupling constant. For later purposes let us also set $s^0=\frac{1}{2}\left(\begin{smallmatrix} 1 & 0 \\ 0 & 1 \end{smallmatrix}\right)$. We then call $H=H_0+H_\text{\sc so}$ the spin-orbit Laplacian on the square lattice. We set $\lambda_\text{\sc so}=1$ as only its ratio with the hopping parameter $t$ is relevant. For the operator $H$, the matrices $T_j$ and $V$ are given by
$$
T_1=t-2\imath s^x\;,
\qquad
T_2=t-2\imath s^y\;,\qquad T_3=V=0\;.
$$
We also have $H_\text{\sc so}({\bf k})=-4\sin(k_1)s^x-4\,\sin(k_2)s^y$ so that
$$
H({\bf k})\;=\;
\begin{pmatrix}
2t(\cos(k_1)+\cos (k_2)) &
2(-\sin(k_1)+\imath\sin (k_2)) \\
2(-\sin(k_1)-\imath\sin (k_2)) & 2t(\cos(k_1)+\cos (k_2))
\end{pmatrix}
\;.
$$
This implies that the energy bands are $E_\pm({\bf k})= 2t(\cos(k_1)+\cos (k_2))\pm 2(\sin^2(k_1)+\sin^2 (k_2))^{\frac{1}{2}}$. Note that even though there are two signs, no gap opens in the spectrum of $H({\bf k})$. Next the partial Fourier transform is given by
$$
\Uu_1\widehat{H}_\text{\sc so}\,\Uu_1^*(k_1)
\;=\;
-\,4\,\sin(k_1)s^x\;+\;2\,
\imath\,(\widehat{S}_2-\widehat{S}_2^*)\otimes s^y
\;.
$$
Thus $\widehat{H}_2(k_1)$ is also a Jacobi matrix with matrix entries ${A}_2(k_1)$ and ${B}_2(k_1)$ given by
$$
{A}_2(k_1)\;=\;
\begin{pmatrix} t &  -\,1
\\
1 & t
\end{pmatrix}
\;,
\qquad
{B}_2(k_1)\;=\;
\begin{pmatrix} 2\,t\cos(k_1) &  -\,2\,\sin(k_1)
\\
-\,2\,\sin(k_1) & 2\,t\cos(k_1)
\end{pmatrix}
\;.
$$
Note that ${A}_2(k_1)$ is invertible so that the transfer matrices $\Tt^E_2(k_1)$ are well-defined.  By studying the unitary $U^E(k_1)$ associated to the contracting directions, it can again readily be verified that there are no edge states when one imposes Dirichlet boundary conditions.

\vspace{.2cm}


\subsection{The Harper model}

\begin{figure}
\begin{center}
{\includegraphics[width=4cm]{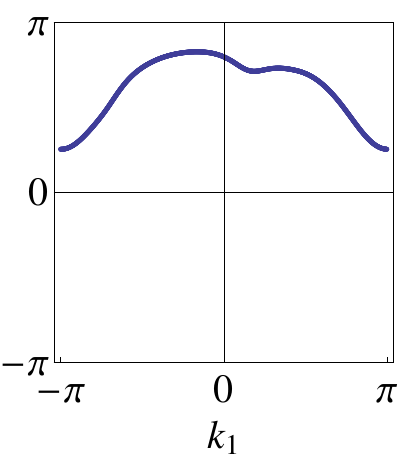}}
\hspace{.3cm}
{\includegraphics[width=4cm]{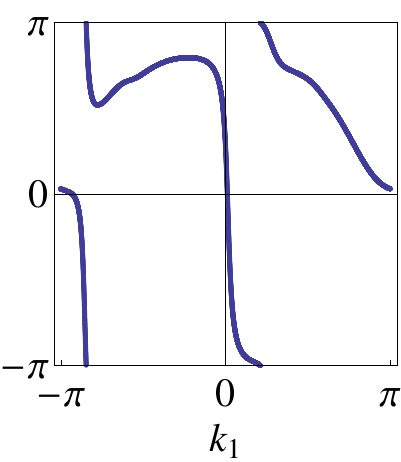}}
\hspace{.3cm}
{\includegraphics[width=4cm]{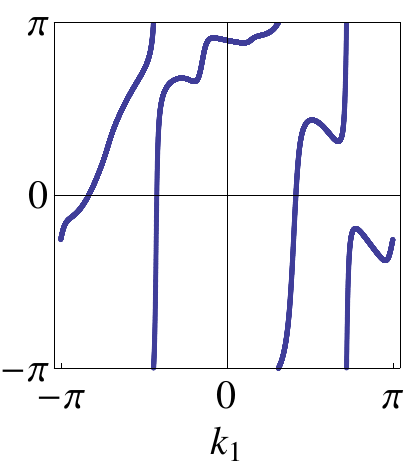}}
\caption{\sl For a Harper model with flux $\varphi=2\pi\,\frac{3}{7}$, the phase of $k_1\in[-\pi,\pi]\mapsto U^E(k_1)$ of the transfer matrix $\Tt_{2,7,1}^E(k_1)$ is plotted for the energies $E=-2.7$, $E=-2.4$ and $E=-1.9$ which lie respectively below the spectrum and in the first and second gap. As expected the edge index of the first is $0$, then for the second and third it is $-2$ and $3$ respectively. Therefore the Chern numbers of the lowest two bands are $-2$ and $5$.}
\label{fig-Harper}
\end{center}
\end{figure}

The Harper model is the magnetic Laplacian on the square lattice with a flux $\varphi$ through each unit cell. Its Hamiltonian in Landau gauge is given by $H=e^{\imath \varphi X_2}S_1+e^{-\imath \varphi X_2}S_1^*+S_2+S_2^*$. The flux $\varphi=2\pi\,\frac{q}{p}$ is chosen to be rational multiple of $2\pi$. There are $p$ transfer matrices in $2$-direction now, as they depend on the site $n$:
$$
\Tt_{2,n}^E(k_1)
\;=\;
\begin{pmatrix}
E-2\cos(k_1+n\varphi) & -1 \\ 1 & 0
\end{pmatrix}
\;.
$$
Then the periodic block consists of the product of $p$ consecutive of these matrices:
$$
\Tt_{2,p,1}^E(k_1)
\;=\;
\Tt_{2,p}^E(k_1)\cdots
\Tt_{2,1}^E(k_1)
\;.
$$
Now the intersection theory is applied to $\Tt_{2,p,1}^E(k_1)$. Numerically, one calculates the eigenvector of this $2\times 2$ matrix lying inside of the unit circle (for an energy $E$ in a gap of $H$ so that all matrices  $\Tt_{2,p,1}^E(k_1)$ have trace smaller than $2$ in absolute value and are indeed hyperbolic) and then one deduces the unitary $U^E(k_1)$ by \eqref{eq-UPhiCalc}. It is just a phase here. This is plotted for one $\varphi$ and three consecutive gaps in Figure~\ref{fig-Harper}.

\subsection{Laplacian on a triangular lattice}

The triangular lattice consists of the nodes $\Gamma={\bf a}_1\ZM+{\bf a}_2\ZM$ spanned by the vectors ${\bf a}_1=\frac{1}{2}(1,\sqrt{3}),{\bf a}_2=\frac{1}{2}(-1,\sqrt{3})$ with edges between next neighbors and supplementary edges in the direction ${\bf a}_3={\bf a}_2-{\bf a}_1$.  Associated to the lattice is the Hilbert space $\ell^2(\Gamma)$ of all sequences of complex numbers indexed by points in $\Gamma$. On $\ell^2(\Gamma)$ act the three shift operators ${\mathcal S}_j$, $j=1,2,3$, along the edges defined by $({\mathcal S}_j\phi)_{{\bf a}}=\phi_{{\bf a}-{\bf a}_j}$ where ${\bf a}\in\Gamma$. Note that $\Ss_3=\Ss_1^*\Ss_2$. Now the next nearest neighbor hopping Hamiltonian on the triangular lattice is
$$
\widetilde{H}_0\;=\;\sum_{i=1,2,3} t_i(\Ss_i+\Ss_i^*)
\;,
$$
with coefficients $t_1,t_2,t_3>0$. Then the dual lattice is $\Gamma^*=\{{\bf b}\in\RM^2\,|\,{\bf b}\cdot{\bf a}\in 2\pi\ZM\;\forall\;{\bf a}\in\Gamma\}$ and the Brillouin zone is $\Bb_\Gamma=\{{\bf k}\in\RM^2\,|\,|{\bf k}| \leq |{\bf k}-{\bf b}|\;\forall\;{\bf b}\in\Gamma^*\}$. It turns out to be a hexagon with volume $\vol(\Bb_\Gamma)=(2\pi)^2/|{\bf a}_1\times {\bf a}_2|$. The $\Gamma$-Fourier transform $\Uu_\Gamma=\ell^2(\Gamma)\to L^2(\Bb_\Gamma,\frac{d{\bf k}}{\vol(\Bb_\Gamma)})$ is defined by
$$
(\Uu_\Gamma\phi)({\bf k})
\;=\;
\sum_{{\bf a}\in\Gamma}\phi_{\bf a} e^{\imath{\bf a}\cdot {\bf k}}
\;.
$$
The triangular lattice has a $\frac{\pi}{3}$ rotational symmetry which it inherits to $\Bb_\Gamma$. The Hamiltonian $\widetilde{H}_0$ has this symmetry if and only if $t_1=t_2=t_3$. If it is not given,  there is also no need to use the above Brillouin zone instead of the homeomorphic torus. We will rather choose to deform $\Gamma$ into a square lattice $\ZM^2$ which will then lead to the framework of Section~\ref{sec-models}. For this purpose let us introduce the $2\times 2$ invertible matrix $A_\Gamma=({\bf a}_1,{\bf a}_2)$. Then $\Gamma=A_\Gamma\ZM^2$ and $\ZM^2=A_\Gamma^{-1}\Gamma$. Associated is the unitary ${\mathcal V}:\ell^2(\Gamma)\to\ell^2(\ZM^2)$ given by $({\mathcal V} \phi)_{{\bf n}}=\phi_{A_\Gamma {\bf n}}$ for ${\bf n}\in\ZM^2$. After this unitary transformation $S_j={\mathcal V}\Ss_j{\mathcal V}^*$ are the translations on $\ell^2(\ZM^2)$ as defined in Section~\ref{sec-models} and the Hamiltanion $H_0={\mathcal V}\widetilde{H}_0{\mathcal V}^*$ is
$$
H_0\;=\;\sum_{i=1,2,3} t_i(S_i+S_i^*)
\;.
$$
It is hence of the form \eqref{eq-genH} with coefficients
$$
T_i=t_i\;,\qquad V=0\;.
$$
Here there is no internal degree of freedom and we have not added a spin so far, so for now $R=r=L=1$. A spin orbit interaction can be added, similar to the square lattice or the honeycomb lattice dealt with below.

\vspace{.2cm}

Now let us first proceed with the analysis of $H_0$. For the calculation of the spectrum, we use
$$
H_0({\bf k})\;=\;
\sum_{i=1,2,3} 2\,t_i\cos(k_i)
\;,
\qquad
k_3=k_2-k_1
\;.
$$
We thus have $\max_{{\bf k}\in\TM^2}H_0({\bf k})=2(t_1+t_2+t_3)$ (realized for ${\bf k}=0$). 
In case $t_j=1$ we also find $\min_{{\bf k}\in\TM^2}H_0({\bf k})=-3$ and the spectrum is $\sigma(H_0)=[-3,6]$.  Next let us introduce $\rho=\rho(k_1)\geq 0$ and $\theta=\theta(k_1)$ by
$$
e^{\imath\theta}
\;=\;
\det(\Tt_2^E(k_1))^{\frac{1}{2}}
\;=\;
\left(\frac{e^{-\imath k_1}t_3+t_2}{e^{\imath k_1}t_3+t_2}\right)^{\frac{1}{2}}
\;,
$$
and $\rho=|e^{-\imath k_1}t_3+t_2|$. Then $e^{-\imath k_1}t_3+t_2=\rho e^{\imath\theta}$. Hence
\begin{eqnarray*}
\Tt_2^E(k_1)
& = &
\begin{pmatrix}
(E-2t_1\cos(k_1))(e^{\imath k_1}t_3+t_2)^{-1} & -(e^{\imath k_1}t_3+t_2)^* \\
(e^{\imath k_1}t_3+t_2)^{-1} & 0
\end{pmatrix}
\\
& = &
e^{\imath\theta}
\begin{pmatrix}
(E-2t_1\cos(k_1))\rho^{-1} & -\rho  \\
\rho^{-1} & 0
\end{pmatrix}
\end{eqnarray*}
The spectrum of $\Tt_2^E(k_1)$ now consists of two eigenvalues
$$
\lambda_\pm(k_1)
\;=\;
e^{\imath\theta}\,\left(\frac{E-2t_1\cos(k_1)}{2\rho}\pm\sqrt{\left(\frac{E-2t_1\cos(k_1)}{2\rho}\right)^2-1}\;\right)
\;.
$$
Let us examine these eigenvalue curves in the situation where $E\not\in\sigma(H)$ so that the eigenvalue curves never intersect $\SM^1$. Due to the symplectic nature of $\Tt_2^E(k_1)$ one of these curves is then within $\SM^1$, the other outside of $\SM^1$. For $t_2>t_3$, neither of these curves makes a loop around the origin (because the factor $e^{\imath\theta}$ does not and the other factor is real of definite sign). For $t_2<t_3$, however, each curve makes a loop around $0$. At the critical value $t_2=t_3$ corresponding to a Hamiltonian with all symmetries of the triangular lattice, one has $e^{\imath \theta}=e^{-\imath \frac{k_1}{2}}$ and $\rho=2\cos(\frac{k_1}{2})$ so that  $\lim_{k_1\uparrow\pi}e^{\imath\theta}=-\imath$, $\lim_{k_1\downarrow-\pi}e^{\imath\theta}=\imath$ and $\lim_{k_1\to\pi}\rho=0$. Thus the inner eigenvalue curve crosses the origin parallel to the imaginary axis, while the outer curve diverges to $-\imath\infty$ and $\imath\infty$ (seen as directions in $\CM$). This are the singularities alluded to in Section~\ref{sec-models}. As for the Laplacian on the square lattice, one can now check that there are no edge states for the triangular lattice.

\subsection{Laplacian on the honeycomb lattice}

It is well-known that the honeycomb lattice can be viewed as a decorated triangular lattice. Indeed, let $\Gamma={\bf a}_1\ZM+{\bf a}_2\ZM$ be the triangular lattice defined in the last section. Then the points $\Gamma\cup (\Gamma-{\bf d})$ with ${\bf d}=(0,1/\sqrt{3})$ constitute the nodes of the honeycomb lattice. From each point there are three edges to the three closest points which for points in $\Gamma$ lie in $\Gamma-{\bf d}$ and visa versa. The honeycomb lattice has $\frac{2\pi}{3}$ rotational symmetry, but again this symmetry will not be necessarily conserved by the operators under study. Next we pass from $\Gamma$ to $\ZM^2$ exactly as in the last section and then treat the states on the sites ${\bf a}\in\Gamma$ and ${\bf a}-{\bf d}$ as an internal degree of freedom over each point of the triangular lattice. Hence here $R=2$ (the spin is only relevant in the next section).  After the unitary transformation ${\mathcal V}$ described above, the Hilbert space is thus $\ell^2(\ZM^2)\otimes\CM^2$. The discrete Laplacian connecting edges of the honeycomb lattice is now in the grading of $\CM^2$ given by
$$
H_0
\;=\;
\begin{pmatrix}
0 & t_1S^*_1+t_2S^*_2+t_3 \\
t_1S_1+t_2S_2+t_3 & 0
\end{pmatrix}
\;.
$$
The fact that it only contains off-diagonal entries reflects that edges only connect nodes of $\Gamma$ to nodes of $\Gamma-{\bf d}$. Again $H_0$ is of the form \eqref{eq-genH} with coefficients
$$
T_1=\begin{pmatrix} 0 & t_1 \\ 0 & 0 \end{pmatrix}\;,
\qquad
T_2=\begin{pmatrix} 0 & t_2 \\ 0 & 0 \end{pmatrix}\;,
\qquad
T_3=0\;,
\qquad
V=\begin{pmatrix}  0 & t_3  \\ t_3 & 0 \end{pmatrix}\;.
$$
Note that here $T_1$ and $T_2$ are not invertible, nor is $T_3S_2^*+T_1$. Thus the transfer operators cannot be defined in this case. The reason is that there are not sufficiently many non-vanishing matrix elements between the lattice sites. A way out is to rather work with two different consecutive transfer operators (from one sublattice to the other), but we will not need to do so because in presence of the spin-orbit hopping terms added in the next section, the invertibility conditions are satisfied and then the transfer operators can again be defined as before.

\vspace{.2cm}
 
Again the operator $H_0$ can be diagonalized by the Fourier transform \eqref{eq-Fourier}:
$$
H_0({\bf k})
\;=\;
\begin{pmatrix}
0 & t_1\e^{-\imath k_1}+t_2\e^{-\imath k_2}+t_3\\
t_1\e^{\imath k_1}+t_2\e^{\imath k_2}+t_3 & 0
\end{pmatrix}\;.
$$
Then there are two energy bands $E_{+}$ and $E_{-}$ given by
$$
E_\pm({\bf k})
\;=\;
\pm\,
|t_1\e^{\imath k_1}+t_2\e^{\imath k_2}+t_3|
\;.
$$
Depending on the respective values of the hopping terms, the bands can overlap or be separated by a gap. In the particular case $t_j=t$ for $j=1,2,3$, the formula for the energy reduces to
$$
E_\pm({\bf k})
\;=\;
\pm\, t\,\sqrt{3+2\cos(k_1-k_2)+2\cos(k_1)+2\cos(k_2)} 
\;.
$$
In this case there are precisely two points $K,K'\in\TM^2$ (both giving energy $0$)
at which the bands touch. These points are called Dirac points (which differ from those in the literature due to our deformation of the Brillouin zone).

\vspace{.2cm}

Again let us consider $\widehat{H}_0$. This corresponds to a half-space of the hexagonal lattice with a zig-zag boundary in the direction ${\bf a}_1$. (A so-called armchair boundary can be obtained by cutting the direction ${\bf a}_3$.) Then $\widehat{H}_{0,2}(k_1)$ acting on $\ell^2(\NM)\otimes\CM^2$ is given by
$$
\widehat{H}_{0,2}(k_1)
\;=\;
\begin{pmatrix}
0 & t_1\e^{-\imath k_1}+t_2\widehat{S}_2^*+t_3\\
t_1\e^{\imath k_1}+t_2\widehat{S}_2+t_3 & 0
\end{pmatrix}\;.
$$
Hence $\widehat{H}_{0,2}(k_1)$ is a Jacobi matrix with $2\times 2$ matrix entries
\begin{equation}
\label{eq-Jacentries}
{A}_2(k_1)
\;=\;
\begin{pmatrix}
0 & t_2
\\
0 & 0
\end{pmatrix}
\;,
\qquad
{B}_2(k_1)
\;=\;
\begin{pmatrix}
0 & t_3+t_1\e^{\imath k_1}
\\
t_3+t_1\e^{-\imath k_1} & 0
\end{pmatrix}
\;.
\end{equation}
In the present case ${A}_2(k_1)$ is not invertible and the transfer matrices cannot be defined, so that the representation as a Jacobi matrix is basically useless. However, the same representation holds also when the spin orbit interaction is included in the next section, and then the entries ${A}_2(k_1)$ will turn out to be invertible for almost all $k_1$. In the present situation without spin-orbit coupling, it is possible to choose yet another representation of $\widehat{H}_{0,2}(k_1)$ resulting in a $2$-periodic Jacobi matrix with complex entries.

\subsection{Hamiltonian on honeycomb lattice with spin-orbit interaction}

Let $H_0$ be the Laplacian on the honeycomb lattice as considered in the last section. Following Kane and Mele \cite{KM}, we now add further translation invariant terms to it, notably the spin-orbit interaction $H_\text{\sc so}$ coupling next-nearest neighbors, a staggered potential $H_\text{\sc st}$ distinguishing the two sublattices, and the Rashba spin-orbit coupling $H_\text{\sc R}$ (which is a nearest neighbor spin-orbit coupling). Hence we add the spin degree of freedom by tensorizing with $\CM^r$ with $r=2s+1$. Then
$$
H_\text{\sc so}
\;= \;
2\,\imath\lambda_\text{\sc so}\sum\limits_{i=1}^3 
\begin{pmatrix} 
t'_{i}(S_i^*-S_i) s^z
& 0
\\
0& -t'_{i}(S_i^*-S_i) s^z
\end{pmatrix}
\;,
\qquad
H_\text{\sc st}
\; = \;
\lambda_\text{\sc st}\begin{pmatrix} 1 & 0 \\ 0 & -1 \end{pmatrix}
\;.
$$
Then for $H_0+H_\text{\sc so}+H_\text{\sc st}$ the coefficient matrices are, for $j=1,2$
$$
T_j=\begin{pmatrix} 2\imath \lambda_\text{\sc so}t'_js^z& t_j \\ 0 & -2\imath \lambda_\text{\sc so}t'_js^z \end{pmatrix}\;,
\qquad
T_3=\begin{pmatrix} 2\imath \lambda_\text{\sc so}t'_3s^z & 0\\ 0 & -2\imath \lambda_\text{\sc so}t'_3s^z \end{pmatrix}\;,
\qquad
V=\begin{pmatrix}  \lambda_\text{\sc st} & t_3  \\ t_3 & -\lambda_\text{\sc st} \end{pmatrix}\;.
$$
Moreover, the Rashba term is with the notations $S'_1=S_1$, $S'_2=S_2$, $S'_3={\bf 1}$,
$$
H_\text{\sc r}
\;= \;
2\,\imath\lambda_\text{\sc r}\sum\limits_{i=1}^3 
\begin{pmatrix} 
0 & t''_{i}(S'_i)^*  ({\bf d}_{i}\times{\bf s})_z
\\
-t''_{i}S'_i  ({\bf d}_{i}\times{\bf s})_z
& 0
\end{pmatrix}
\;,
$$
where
$$
{\bf d}_{1}=\frac{{\bf a}_1-{\bf d}}{\|{\bf a}_1-{\bf d}\|}\;,
\qquad
{\bf d}_{2}=\frac{{\bf a}_2-{\bf d}}{\|{\bf a}_2-{\bf d}\|}\;,
\qquad
{\bf d}_{3}=-\,\frac{{\bf d}}{\|{\bf d}\|}\;.
$$
If $H_\text{\sc r}=0$, then $H=H_0+H_\text{\sc so}+H_\text{\sc st}$ which commutes with $s^z$. Thus in the grading given by the spectral subspaces of $s^z$, the Hamiltonian is diagonal in this situation and can be written as $H=H_+\oplus H_-$ where $H_+$ is restriction of $H$ to the spin up subspace. Similarly, also the transfer operators are diagonal given by $\Tt_j^E=\Tt_{j,+}^E\oplus\Tt_{j,-}^E$. Both $H_\pm$ and $\Tt^E_{j,\pm}$ are obtained simply by replacing $2s^z$ by $\pm 1$.

\vspace{.2cm}

After Fourier transform, the components of the Hamiltonian become
\begin{eqnarray*}
H_\text{\sc so}({\bf k})
& = &
\begin{pmatrix}
4\lambda_\text{\sc so}\sum\limits_{j=1}^3 t'_{j}\sin(k_j)s^z & 0 \\
0 & -4\lambda_\text{\sc so}\sum\limits_{j=1}^3 t'_{j}\sin(k_j)s^z
\end{pmatrix}
\;,
\\
H_\text{\sc st}({\bf k})
& = &
\begin{pmatrix}
\lambda_\text{\sc st} & 0 \\
0 & -\lambda_\text{\sc st}
\end{pmatrix}\otimes {\bf 1}
\;,
\\
H_\text{\sc r}({\bf k})
& = &
\imath\,\lambda_\text{\sc r}\begin{pmatrix} 0 & t''_1\e^{-\imath k_1}+t''_2\e^{-\imath k_2}-2t''_3 \\ -t''_1\e^{\imath k_1}-t''_2\e^{-\imath k_2}+2t''_3 & 0 \end{pmatrix}\otimes s^x \\
& & +\,\imath\,\lambda_\text{\sc r} \begin{pmatrix} 0  & -\sqrt{3}(t''_1\e^{-\imath k_1}-t''_2\e^{-\imath k_2}) \\ \sqrt{3}(t''_1\e^{\imath k_1}-t''_2\e^{\imath k_2}) & 0 \end{pmatrix}\otimes s^y
\;.
\end{eqnarray*}
In order to simplify these expressions somewhat let us choose $t'_{j}=t''_j=1$ and  also use the Pauli matrices $(\sigma_x,\sigma_y,\sigma_z)=2\,{\bf s}$ for the pseudo-spin degree of freedom. Then
\begin{eqnarray*}
H_0({\bf k})
& = &
t(\cos (k_1)+\cos(k_2)+1 )\sigma^x\otimes {\bf 1}+t(\sin(k_1)+\sin (k_2))\sigma^y\otimes {\bf 1}
\\
H_\text{\sc so}({\bf k})
& = &
4\,\lambda_\text{\sc so}\sum_{j=1,2,3}\sin (k_j)\sigma^z\otimes s^z
\;,
\\
H_\text{\sc st}({\bf k})
& = &
\lambda_\text{\sc st}\,\sigma^z\otimes{\bf 1}
\;,
\\
H_\text{\sc r}({\bf k})
& = &
\lambda_\text{\sc r}\,\Big[(-\cos(k_1)-\cos (k_2)+2)\sigma^y\otimes s^x+(\sin (k_1)+\sin (k_2)) \sigma^x\otimes s^x
\\
& & \;\;\;\;\;+\,\sqrt{3}(\cos (k_1)-\cos (k_2)) \sigma^y\otimes s^y+\sqrt{3}(-\sin (k_1)+\sin (k_2))\sigma^x\otimes s^y\Big]
\;.
\end{eqnarray*}
Hence setting $H$ the sum of the four terms, we can expand $H({\bf k})$ in the basis given by $5$ tensor products
$\Lambda^n$ of Pauli matrices as given in the table below and their $10$ commutators $\Lambda^{nm}=[\Lambda^n,\Lambda^m]/(2i)$:
$$
H({\bf k})
\;=\;
\sum_{n=1}^5\alpha_n({\bf k})\Lambda^n+\sum_{n<m=1}^5\alpha_{nm}({\bf k})\Lambda^{nm}\;.
$$
The following table then gives the non-vanishing coefficients of $H({\bf k})$ w.r.t. this basis.

\vspace{.2cm}

\small\begin{center}\makebox[0pt]{
\(\begin{array}{||l@{\rule{0pt}{10pt}}|l||l|l||}\hhline{|t:==:t:==:t|}
\Lambda^1=\sigma^x\otimes {\bf 1} & \alpha_1=t(\cos (k_1)+\cos( k_2)+1) & \Lambda^{12}=-\sigma^y\otimes {\bf 1} & \alpha_{12}=-t(\sin (k_1)+\sin (k_2)) \\\hhline{||--||--||} \Lambda^2=\sigma^z\otimes {\bf 1} & \alpha_2=\lambda_\text{\sc st} & \Lambda^{15}=\sigma^z\otimes s^z & \alpha_{15}=4\lambda_\text{\sc so}(\sin (k_1)+\sin (k_2)+\sin (k_3)) \\\hhline{||--||--||}
\Lambda^3=\sigma^y\otimes s^x & \alpha_3=\lambda_\text{\sc r}(2-\cos (k_1)-\cos (k_2)) & \Lambda^{23}=-\sigma^x\otimes s^x & \alpha_{23}=-\lambda_\text{\sc r}(\sin (k_1)+\sin (k_2) \\\hhline{||--||--||}
\Lambda^4=\sigma^y\otimes s^y & \alpha_4=\lambda_\text{\sc r}\sqrt{3}(\cos (k_1)-\cos (k_2)) & \Lambda^{24}=-\sigma^x\otimes s^y & \alpha_{24}=\lambda_\text{\sc r}\sqrt{3}(\sin (k_1)-\sin (k_2)) \\\hhline{||--||--||}
\Lambda^5=\sigma^y\otimes s^z & \alpha_5=0 & & \\\hhline{|b:==:b:==:b|}
\end{array}\)}\end{center}\normalsize

\vspace{.2cm}

\begin{figure}
\begin{center}
\includegraphics[width=4cm]{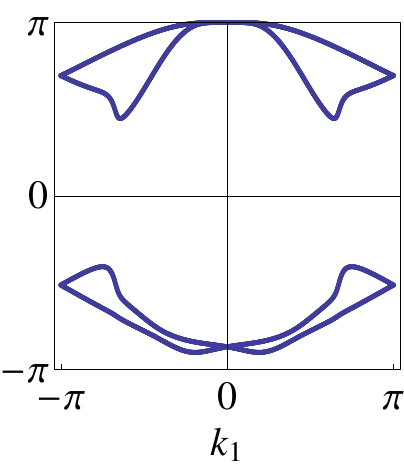}
\hspace{.3cm}
\includegraphics[width=4cm]{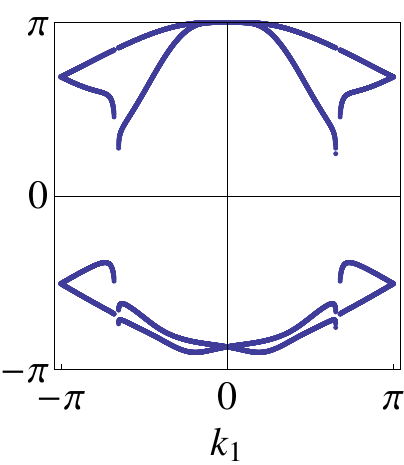}
\hspace{.3cm}
\includegraphics[width=4cm]{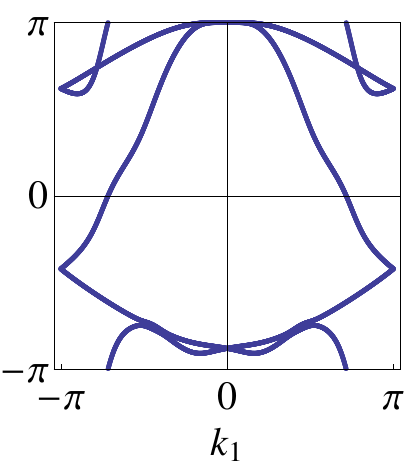}
\caption{\sl For the Kane-Mele model, the spectrum of the unitaries $k_1\in[-\pi,\pi]\mapsto U^E(k_1)$ calculated from {\rm \eqref{eq-UPhiCalc}} is plotted for three different values of the parameters. The energy is always $E=0$, furthermore $t_1=t_2=t_3=t'_1=t'_2=1$ and $t'_3=0.9$ as well as $\lambda_\text{\sc r}=0.3$ and $\lambda_\text{\sc st}=0.45$. In the left graph $\lambda_\text{\sc so}=0.89$ and the edge $\ZM_2$-index is trivial, while in the right graph, $\lambda_\text{\sc so}=1$ and the index is non-trivial. In the middle graph $\lambda_\text{\sc so}=0.9$, a point that is close to the transition between the phases. 
}
\label{fig-KaneMele}
\end{center}
\end{figure}

Next let us restrict $H$ to the half-space. Proceeding as in the last section, the fibered operators $\widehat{H}_2(k_1)$ can be written as a Jacobi matrix as in \eqref{eq-Hqdef}, but the matrix entry ${B}_2(k_1)$ is now of size $4\times 4$ and given by
$$
\begin{pmatrix}
\lambda_\text{\sc st} \,{\bf 1}-
4\,\lambda_\text{\sc so} \sin(k_1)s^z
& \!\!\!\!\!\! t(\e^{\imath k_1}+1)\,{\bf 1}+\imath\,
\lambda_\text{\sc r} (\e^{-\imath k_1}-2)s^x-\imath\sqrt{3}
\lambda_\text{\sc r}\e^{-\imath k_1}s^y
\\
t(\e^{-\imath k_1}+1)\,{\bf 1}-\imath
\lambda_\text{\sc r} (\e^{\imath k_1}-2)s^x+\imath\sqrt{3}
\lambda_\text{\sc r} \e^{\imath k_1}s^y &
-\lambda_\text{\sc st} \,{\bf 1}+4\,
\lambda_\text{\sc so}\sin(k_1)s^z
\end{pmatrix}
,
$$
and
$$
{A}_2(k_1)
\,=\,
\begin{pmatrix}
2\imath
\lambda_\text{\sc so}(1- \e^{-\imath k_1})s^z & t
+\imath
\lambda_\text{\sc r} s^x
+
\imath\sqrt{3}
\lambda_\text{\sc r} s^y
\\
0 & -2\imath
\lambda_\text{\sc so}(1- \e^{-\imath k_1})s^z
\end{pmatrix}\;.
$$
The main difference with the case without spin-orbit interaction is that for almost all $k_1$ the matrices ${A}_2(k_1)$ are invertible so that the transfer matrices
${\Tt}_2^z(k_1)$ can again be defined. From this the unitary $U^E(k_1)$ and its spectrum can be calculated numerically using \eqref{eq-UPhiCalc}. Some examples are plotted in Figure~\ref{fig-KaneMele}.

\appendix

\section{Periodic Jacobi matrices with matrix entries}
\label{app-periodic}

In this appendix we study periodic Jacobi matrices $H$ acting on the Hilbert space $\Hh=\ell^2(\ZM,\CM^L)$ as well as their half-line restrictions $\widehat{H}$ acting on $\widehat{\Hh}=\ell^2(\NM,\CM^L)$. The operators $H$ are of the form
$$
H
\;=\;
A\,S^*\,+B\,
+\,A^*S
\;,
$$
where $S$ is the unitary left shift and $A$ and $B$ are $L\times L$ matrices with $B=B^*$. Then $\widehat{H}$ is defined using the one-sided shift $\widehat{S}$ on $\widehat{\Hh}$ which is only a partial isometry. With standard modifications all the proofs below also transpose the situation of $p$-periodic operator instead of a $1$-periodic operator. 

\subsection{Spectral analysis of the two-sided operator}
\label{app-twosidedperiodic}

The Hamiltonian $H$ is diagonalized by the Fourier transform $\Uu:\ell^2(\ZM,\CM^L)\to L^2(\TM^1,\frac{dz}{2\pi})\otimes \CM^L$ defined  $(\Uu\psi)(z)=\sum_{n\in\ZM}\psi_{n} z^n$, similar as in \eqref{eq-Fourier}. Then $\Uu H\Uu^*=\int_{\TM^1}\frac{dz}{2\pi}\,H(z)$ with
$$
H(z)\;=\;A\,z^{-1}\,+B\,+\,A^*\,z
\;.
$$
Note that for $z\in \TM^1\subset\CM$, the matrix $H(z)$ is self-adjoint. Somewhat abusing notations, we also write $H(q)$ for $H(e^{\imath q})$. By analytic perturbation theory, $H(q)$ has $L$ eigenvalues $E_l(q)$, $l=1,\ldots,L$, which at level crossings can be chosen to be analytic. As $q$ runs through $(-\pi,\pi]$ each eigenvalue leads to an energy band of $H$. The spectrum of $H$ is absolutely continuous if none of the eigenvalues is constant in $q$. If such a constant energy occurs, then one also speaks of a flat band. It leads to a Dirac peak in the density of states. The proof of the following result is inspired by the arguments in Section~5 of \cite{Kuc}.

\begin{proposi}
\label{prop-Ainvertible}
If $A$ is invertible, there are no flat bands so that $H$ has absolutely continuous spectrum. If there is a flat band at energy $E$, then $H$ has compactly supported eigenstates at $E$. 
\end{proposi}

\noindent {\bf Proof.} Suppose that there is a flat band with energy $E$. Then $\det(H(z)-E)=0$ for all $z\in\TM^1$. Now $H(z)$ is a Laurent polynomial in $z$ so that $\det(H(z)-E)$ is also a Laurent polynomial in $z$. Hence $\det(H(z)-E)=z^{-L}P(z)$ with some polynomial $P$. Now $\det(H(z)-E)=0$ is equivalent to $P(z)=0$. But a polynomial can only vanish on $\TM^1$ if it vanishes everywhere so that $P=0$. But then $\det(H(z)-E)=0$ for all $z\in\CM$. This, however, is not possible because $\det(H(z)-E)=z^L\det(A^*)+\Oo(z^{L-1})$, so for $z$ sufficiently large the determinant $\det(H(z)-E)$ does not vanish since $\det(A)\not =0$. 

\vspace{.2cm}

For the second claim, let us now suppose that $E$ is the energy of a flat band. Then $H(z)-E$ has a non-trivial kernel for every $z\in\TM^1$. Calculating a vector $v(z)$ in this kernel by the standard Gauss algorithm, we see that it can also be chosen to be a Laurent polynomial in $z$. The Fourier transform $\Uu^*v$ of its restriction to $\TM^1$ is therefore compactly supported. It can be shown that translations of the compactly supported state actually span the full eigenspace of $E$ \cite{Kuc}.
\hfill $\Box$

\vspace{.2cm}

\noindent {\bf Remark} Let us provide another proof that $H$ has no point spectrum if $A$ is invertible. This argument uses transfer matrices and it will be useful to have it in mind when studying edge states below. First we recall that the formal solutions $\psi=(\psi_n)_{n\in\ZM}$ of the Schr\"odinger equation ${H}\psi=E\psi$ at energy $E$ can be analyzed using the transfer matrices (well-defined because $A$ is invertible)
\begin{equation}
\label{eq-transdef}
\Tt^E\;=\;
\begin{pmatrix}
(E\,{\bf 1}-B)A^{-1}
& -\,A^*
\\
A^{-1} & 0
\end{pmatrix}\;,
\end{equation}
namely for $n\in\ZM$
\begin{equation}
\label{eq-transrel}
\begin{pmatrix}
A\psi_{n+1} \\ \psi_{n}
\end{pmatrix}
\;=\;
\Tt^E\,
\begin{pmatrix}
A\psi_n \\ \psi_{n-1}
\end{pmatrix}\;.
\end{equation}
Now if $E$ is an eigenvalue, the corresponding eigenstate $\psi$ has to be square-integrable. Square-summability at $+\infty$  implies that $\psi_0,\psi_1\in\CM^L$ have to be such that the vector $\binom{A\psi_1}{\psi_0}$ lies in a strictly contracting direction of $\Tt^E$. But then it lies in a strictly expanding direction of $(\Tt^E)^{-1}$ and hence $\psi$ is not square-summable at $-\infty$. Let us also point out that $E\in\sigma(H)$ is equivalent to the fact that the transfer matrix $\Tt^E$ has an eigenvalue on the unit circle.
\hfill $\diamond$

\vspace{.2cm}

Next let us show by an example that there may very well be flat bands if $A$ is singular.

\vspace{.2cm}

\noindent {\bf Example.} Suppose that $V=\ker(A)$ is non-trivial so that $M=\dim(V^\perp)$ is strictly less than $L$. Then split $\CM^L=V\oplus V^\perp$. In this decomposition,
$$
A\;=\;
\begin{pmatrix}
0 & a_2 \\
0 & a_3
\end{pmatrix}
\;,
\qquad
B\;=\;
\begin{pmatrix}
b_1 & b_2 \\
b_2^* & b_3
\end{pmatrix}
\;.
$$
Thus
$$
\det(H(z)-E)
\;=\;
\det
\begin{pmatrix}
b_1-E & b_2+a_2 z^{-1} \\
b_2^* + a^*_2 z & b_3-E+a_3z^{-1}+a_3^* z
\end{pmatrix}
\;.
$$
If $a_2=0$, then $a_3$ is invertible and, moreover, $\ker(A)=\ker(A^*)$. We first consider this case. Then one has $\det(H(z)-E)=\det(b_1-E)\det(a_3^*)z^M+\Oo(z^{M-1})$ so that by the above argument, there cannot be a flat band at any energy not in the spectrum of the self-adjoint matrix $b_1$. However, there may be a flat band at an energy $E$ in the spectrum of $b_1$. For instance, if $b_2=0$ then clearly any eigenvector $v$ of the self-adjoint matrix $b_1$ with eigenvalue $E$ leads to an infinite number of eigenstates $\psi= \binom{v}{0}\,\delta_{n}$, $n\in\ZM$, of $H$ with eigenvalue $E$. Each of these states is localized at one site $n$.  Now, if $b_2$ does not vanish, but the same vector $v$ is at least in the kernel of $b_2^*$, then one still has a compactly supported eigenstate of $H$.  We suspect that if $a_2$ has maximal rank such that $\ker(A)\cap\ker(A^*)=\{0\}$ there are generically no such eigenstates, but were not able to prove this. Nevertheless, in concrete situations the above proof provides an efficient technique to exclude flat bands: just check that $\det(H(z)-E)$ is not identically equal to $0$ by studying the large or small $z$ limits.

\vspace{.2cm}

\subsection{Spectral analysis of the one-sided operator}
\label{app-onesidedperiodic}

Now let us study the half-space operator $\widehat{H}$. As $\widehat{H}\oplus\widehat{H}$ differs from $H$ only be a finite rank perturbation (the hopping term from $|0\rangle$ to $|-1\rangle$), the essential spectrum of $\widehat{H}$ coincides with that of $H$ and is given by the Bloch bands. We are interested in calculating the new point spectrum, that is the scattering states of the pair $(H,\widehat{H})$. These bound states are related to edge state channels in Section~\ref{sec-edgespec}. For that purpose, let us calculate the solutions of $\widehat{H}\psi=E\psi$ again using the transfer matrices just as in \eqref{eq-transrel} using the same matrices \eqref{eq-transdef}, but now $n$ only runs through $\NM$ and the state $\psi$ has to satisfy the Dirichlet boundary condition $\psi_{-1}=0$. By analyzing the square-summability of the solution at $+\infty$ just as in the remark above, one sees that there is a bound state at $E$ if and only if the contracting directions of $\Tt^E$ contain a vector that satisfies the Dirichlet boundary condition. More precisely, the dimension of the intersection of the plane of contracting directions of $\Tt^E$ in $\CM^{2L}$ with the plane $\binom{\one}{0}$ of Dirichlet boundary conditions is equal to the multiplicity of $E$ as eigenvalue of $\widehat{H}$. In terms of the vector $\begin{pmatrix} A\psi_0 \\ 0 \end{pmatrix}$ in this intersection, the corresponding bound state $\psi$ of ${H}$   is then given by
$$
\psi_n
\;=\;
\begin{pmatrix} 0 \\ \one \end{pmatrix}^*
(\Tt^E)^{n+1}\begin{pmatrix} A\psi_0 \\ 0 \end{pmatrix}\;.
$$

Now the  focus will be on possible bound state energies $E$ lying in a gap of $H$ (and hence not possible embedded eigenvalues). For such a real energy $E\not\in\sigma(H)$ the $\Jj$-unitary transfer matrix $\Tt^E$ is hyperbolic (no eigenvalue on the unit circle). The following result shows that the spectral projection on all eigenvalues of $\Tt^E$ inside the unit disc span a $\Jj$-Lagrangian subspace of $\CM^{2L}$, namely a maximal subspace on which the form $\Jj$ vanishes. Here $\Jj$ is defined in \eqref{eq-symplectictransferop}.

\begin{proposi}
\label{prop-contracting}
For any $\Jj$-unitary $\Tt$ the span of all generalized eigenspaces with eigenvalues of modulus strictly less than $1$ is isotropic.
\end{proposi}

\noindent {\bf Proof.} Let us show the following more general statement which directly implies the proposition. Let $z$ and $z'$ be two {\rm (}possibly equal{\rm )} discrete eigenvalues of a $\Jj$-unitary operator $\Tt$ and denote the associated generalized eigenspaces {\rm (}possibly reducible{\rm )} by $\Ee_z$ and $\Ee_{z'}$. If $z \not = \overline{z'}^{-1}$, then $\Ee_z$ and $\Ee_{z'}$ are $\Jj$-orthogonal. First let $v\in\Ee_z$ and $v'\in\Ee_{z'}$ be eigenvectors. Then
$$
v^*\Jj v'
\;=\;
\frac{1}{\overline{z}\,z'}\;
(\Tt v)^*\Jj (\Tt v')
\;=\;
\frac{1}{\overline{z}\,z'}\;
v^*\Jj v'
\;.
$$
From this follows indeed $v^*\Jj v'=0$ as long as $\overline{z}\,z'\not = 1$. Next recall that $\Ee_z$ is spanned by a finite chain of linearly independent generalized eigenvectors $v_k$, $k=1,\ldots,p$, satisfying $\Tt v_k=z v_k+v_{k-1}$ for $k\geq 2$ and $\Tt v_1=z v_1$. Hence $v_1$ is an eigenvector. Similarly $\Ee_{z'}$ is spanned by chains of $v'_k$, $k=1,\ldots,q$. For each pair of such chains, the above already shows $v_1^*\Jj v'_1=0$. Next for $v'_2$ one has
$$
v_1^*\Jj v'_2
\;=\;
\frac{1}{z'}\;
v_1^*\Jj (\Tt v'_2-v'_1)
\;=\;
\frac{1}{z'}\;
v_1^*\Jj \Tt v'_2
\;=\;
\frac{1}{\overline{z}\,z'}\;
v_1^*\Jj v'_2
\;,
$$
which implies again $v_1^*\Jj v'_2=0$. Iteratively now follows $v_1^*\Jj v'_k$ for all $k=1,\ldots,q$. Then a similar iteration over the $v_k$ completes the proof.
\hfill $\Box$

\vspace{.2cm}

Proposition~\ref{prop-contracting} will be applied to the transfer matrix $\Tt^E$. Then $E$ is an eigenvalue of $\widehat{H}$ exactly if the $\Jj$-Lagrangian plane of contracting directions of $\Tt^E$ has a non-trivial intersection with the plane of Dirichlet boundary conditions which also form a Lagrangian plane. One is thus led to calculate the dimension of the intersection  of two Lagrangian planes. This is conveniently done by intersection theory. We briefly recall the basic facts needed here, more details can be found {\it e.g.} in \cite{SB}.

\vspace{.2cm}

First let us recall the definition of the stereographic projection of a Lagrangian plane. Let the plane be represented by a $2L\times L$ matrix $\Phi$ of full rank satisfying $\Phi^*\Jj\Phi=0$. If $\Phi=\left(\begin{smallmatrix} a \\ b \end{smallmatrix}\right)$ is decomposed into two $L\times L$ blocks, then stereographic projection is defined by
\begin{equation}
\label{eq-stereographic}
\Pi(\Phi)\;=\;(a-\imath b)(a+\imath b)^{-1}
\;.
\end{equation}
It can then be checked \cite{SB} that the appearing inverse is well-defined and that $\Pi(\Phi)$ is in the unitary group U$(L)$ for Lagrangian $\Phi$ is Lagrangian. Moreover, $\Pi$ establishes a diffeomorphism between the Grassmannian of hermitian symplectic Lagrangian planes and U$(L)$. The main use of the unitary is that the multiplicity of $1$ as eigenvalue the unitary $\Pi(\Phi)$ is equal to the dimension of the intersection of $\Phi$ with the reference Lagrangian plane $\binom{\one}{0}$.  Intersections with other planes can readily be studied because the $\Jj$-unitaries act transitively, but this will not be relevant here.

\vspace{.2cm}

In our application, let $\Phi^E$ be a $2L\times L$ matrix built out of  linear independent vector in the generalized eigenspaces of $\Tt^E$ with eigenvalues of modulus strictly less than $1$. As already pointed out, $\Phi^E$ is then Lagrangian by Proposition~\ref{prop-contracting}  as long as $E$ is real and in a gap of $H$. Now set
$$
U^E\;=\;\Pi(\Phi^E)\;.
$$
Then: 
$$
\mbox{ multiplicity of } 1\mbox{ as eigenvalue of } U^E
\;=\;
\mbox{multiplicity of } E \mbox{ as eigenvalue of } \widehat{H}
\;.
$$ 

\vspace{.2cm}

Next let us give another formula for $U^E$ which is connected to the Weyl-Titchmarch theory of the half-sided Jacobi matrix $\widehat{H}$ (this is developed in detail in \cite{SB2}). Its Green matrix at $E\not\in\sigma(\widehat{H})$ is
$$
\widehat{G}^E
\;=\;
\pi_0^*(\widehat{H}-E)^{-1}\pi_0
\;.
$$
Here $\pi_n:\CM^{L}\to\ell^2(\NM,\CM^{L})$ is the partial isometry on the $n$th site. Hence $\widehat{G}^E$ is an $L\times L$ matrix, which in fact is a Herglotz function of $E$.  The basic fact is now that the plane
$$
\Phi^E
\;=\;
\left(
\begin{array}{c}
-\widehat{G}^E
\\
\one
\end{array}
\right)
\;,
$$
contains all square-integrable solutions of the Sch\"odinger equation $\widehat{H}\psi=E\psi$, but with arbitrary boundary conditions at $0$. Stated differently, the formal solutions defined with the transfer matrices
$$
\psi_n
\;=\;
\left(
\begin{array}{c}
0
\\
\one
\end{array}
\right)^*
(\Tt^E)^n \left(
\begin{array}{c}
-\widehat{G}^E
\\
\one
\end{array}
\right)
v
\;,
\qquad
n\in\NM\;,
$$
are square integrable in $n$ for any $v\in\CM^L$ (for $E$ outside of the spectrum of ${H}$). Therefore, the plane spanned by $\Phi^E=\binom{\widehat{G}^E}{-\one}$ coincides with the plane of contracting directions of $\Tt^E$. Thus plugging this into the stereographic projection, we obtain
\begin{equation}
\label{eq-UGreen}
U^E
\;=\;
(\widehat{G}^E+\imath\,\one)(\widehat{G}^E-\imath\,\one)^{-1}
\;=\;
\left(\one+\imath\,(\widehat{G}^E)^{-1}\right)
\left(\one-\imath\,(\widehat{G}^E)^{-1}\right)^{-1}
\;.
\end{equation}
If $E$ is real, then also this formula implies that $U^E$ is unitary.  Furthermore, we have the following

\begin{theo}
\label{theo-rotprop}
For $E\in \RM/\sigma(H)$, the matrix $U^E$ is unitary and differentiable in $E$. One has 
$$
\frac{1}{\imath}\; (U^E)^*\partial_E U^E\;<\;0\;,
$$
so that the eigenvalues of $U^E$ rotate in the negative sense as a function of real energy.
Furthermore, the multiplicity of $E$ as isolated eigenvalue of $\widehat{H}$ is equal to the multiplicity of $1$ as eigenvalue of $U^E$. 
\end{theo}

\noindent {\bf Proof.} The first and last statements were already proved above. Now let us prove the positivity. We begin from \eqref{eq-UGreen}. Because $G=\widehat{G}^E$ is self-adjoint (for real $E$), we have
$$
(U^E)^*\;=\;(U^E)^{-1}\;=\;(G-\imath\,\one)(G+\imath\,\one)^{-1}
\;=\;\left((G-\imath\,\one)^*\right)^{-1}(G+\imath\,\one)^*\;.
$$
Hence
\begin{eqnarray*}
\frac{1}{\imath}\; (U^E)^*\partial_E U^E
& = &
\frac{1}{\imath}\; (U^E)^* 
\Bigl[\partial_EG\,(G-\imath\,\one)^{-1}
-U^E\partial_EG\,(G-\imath\,\one)^{-1}\Bigr]
\\
& = &
\left((G-\imath\,\one)^{-1}\right)^*
\frac{1}{\imath}\;
\Bigl[
(G+\imath\,\one)^*\partial_EG
-
(G-\imath\,\one)^*
\partial_EG
\Bigr]
(G-\imath\,\one)^{-1}
\\
& = & 
-\,2\,\left((G-\imath\,\one)^{-1}\right)^*
\,\partial_EG
\,(G-\imath\,\one)^{-1}
\;.
\end{eqnarray*}
But
$$
\partial_EG
\;=\;
\pi_0^*(E-\widehat{H})^{-2}\pi_0
\;> 0\;,
$$
completing the proof.

\vspace{.1cm}

Let us provide a second proof for the last statement. First of all, the range of $\pi_1$ is a cyclic subspace for the one-sided Jacobi matrix $\widehat{H}$. Therefore, the matrix valued measure associated to the Herglotz function $z\mapsto\widehat{G}^z$ by the Herglotz representation theorem dominates all spectral measures of $\widehat{H}$. In particular, $E$ is an eigenvalue of $\widehat{H}$ if and only if $\widehat{G}^z$ has a pole at $z=E$.  More precisely, the multiplicities $\nu$ of the eigenvalue and pole are equal. As the eigenvalue is isolated, it follows that $z\mapsto\widehat{G}^z$ is analytic in a pointed neighborhood of $E$ (namely, a ball centered at $E$ with $E$ taken out). Moreover, $\widehat{G}^z$ has no zero in this pointed neighborhood (if it is chosen sufficiently small). Therefore $z\mapsto(\widehat{G}^z)^{-1}$ is well-defined, analytic and can be extended to $E$ because the singularity is removable there. Hence the kernel of $(\widehat{G}^E)^{-1}$ is a subspace of dimension $\nu$. On this subspace $U^E$ acts as the identity as shows the second formula in \eqref{eq-UGreen}.
\hfill $\Box$

\vspace{.2cm}


\end{document}